\documentclass
[aps,prc,twocolumn,showpacs,showkeys,amsmath,floatfix,superscriptaddress]%linenumbers
{revtex4-1}
\usepackage{graphicx}
\usepackage{mathptmx}
\usepackage{dcolumn}
\usepackage{ulem}
\usepackage{bm}
\usepackage{hyperref}
\usepackage{amssymb}
\usepackage{txfonts}
\usepackage{dcolumn}
\usepackage{slashed}
\usepackage{comment}
\usepackage{mathrsfs}
\usepackage{multirow}
\usepackage{float}
\usepackage{color}
\usepackage{indentfirst}
\usepackage[section]{placeins}
\newcommand{\delete}{\bgroup\markoverwith{\textcolor{red}{\rule[0.5ex]{2pt}{1pt}}}\ULon}

\allowdisplaybreaks

\begin{document}
\title{Pairing phase transition: A Finite-Temperature Relativistic Hartree-Fock-Bogoliubov study}
\author{Jia Jie Li}
\affiliation{Institut de Physique Nucl\'{e}aire, IN2P3-CNRS, Universit\'{e} Paris-Sud, F-91406 Orsay, France}
\affiliation{Institut de Physique Nucl\'{e}aire de Lyon, IN2P3-CNRS, Universit\'{e} de Lyon, F-69622 Villeurbanne, France}
\affiliation{School of Nuclear Science and Technology, Lanzhou University, Lanzhou 730000, China}
\author{J\'{e}r\^{o}me Margueron}
\affiliation{Institut de Physique Nucl\'{e}aire de Lyon, IN2P3-CNRS, Universit\'{e} de Lyon, F-69622 Villeurbanne, France}
\author{Wen Hui Long}
\affiliation{School of Nuclear Science and Technology, Lanzhou University, Lanzhou 730000, China}
\author{Nguyen Van Giai}
\affiliation{Institut de Physique Nucl\'{e}aire, IN2P3-CNRS, Universit\'{e} Paris-Sud, F-91406 Orsay, France}

\begin{abstract}
\begin{description}
\item[Background]
The relativistic Hartree-Fock-Bogoliubov (RHFB) theory has recently been developed and it provides a unified
and highly predictive description of both nuclear mean field and pairing correlations.
Ground state properties of finite nuclei can accurately be reproduced without neglecting exchange (Fock) contributions.
\item[Purpose]
Finite-temperature RHFB (FT-RHFB) theory has not yet been developed, leaving yet unknown its predictions for
phase transitions and thermal excitations in both stable and weakly bound nuclei.
\item[Method]
FT-RHFB equations are solved in a Dirac Woods-Saxon (DWS) basis considering two kinds of pairing interactions: finite or zero range.
Such a model is appropriate for describing stable as well as loosely bound nuclei since the basis states have correct
asymptotic behaviour for large spatial distributions.
\item[Results]
Systematic FT-RH(F)B calculations are performed for several semi-magic isotopic/isotonic chains comparing the
predictions of a large number of Lagrangians, among which are PKA1, PKO1 and DD-ME2.
It is found that the critical temperature for a pairing transition generally follows the rule
$T_c = 0.60\Delta(0)$ for a finite-range pairing force and $T_c = 0.57\Delta(0)$ for a contact pairing force,
where $\Delta(0)$ is the pairing gap at zero temperature.
Two types of pairing persistence are analysed: type I pairing persistence occurs in closed subshell nuclei while
type II pairing persistence can occur in loosely bound nuclei strongly coupled to the continuum states.
\item[Conclusions]
This first FT-RHFB calculation shows very interesting features of the pairing correlations at finite temperature and in finite systems such as pairing re-entrance and pairing persistence.
\end{description}
\end{abstract}


\maketitle

\section{INTRODUCTION}

The thermodynamical properties of excited nuclei have drawn over the last decades a renewed attention due to the advanced accurate measurements of level densities at low excitation energies~\cite{Melby1999, Agvaanluvsan2004, Kalmykov2007, Toft2010, Guttormsen2014}. Pairing correlations play an essential role in many Fermion systems and have thus a strong influence on nuclear structure at low excitation energies~\cite{Balian1999, Liu2001, Guttormsen2001, Dean2003, Agvaanluvsan2009}. Pairing correlations in finite systems such as nuclei or Wigner-Seitz cells, and in infinite ones such as in neutron star matter, may exhibit different behaviors reflected in the specific heat and the level density~\cite{Sandulescu2004, Fortin2010, Pastore2011, Margueron2012a, Pastore2013}. Moreover, the phase transition is a complex and rich phenomenon, where pairing re-entrance in asymmetric matter, in odd-nuclei, rotating nuclei, and even in doubly magic nuclei close to the drip line may occur~\cite{Sedrakian1997, Balian1999, Hung2011, Margueron2012a}. The interplay between temperature and shell effects in superfluid systems, giving rise to re-entrance or its opposite phenomenon --- suppression, still remains to be studied.

The competition between temperature and pairing correlations in nuclei at low excitation energies has been studied for several decades, with the pioneering works based on finite-temperature BCS (FT-BCS) theory~\cite{Sano1963} and finite-temperature Hartree-Fock-Bogoliubov (FT-HFB) theory~\cite{Goodman1981}. It was predicted that the critical temperature $T_c$ for pair correlation quenching could be expressed, as in uniform matter, as a function of the average pairing gap at zero temperature $\Delta(0)$ following the relations: $T_c = 0.57\Delta(0)$ for a constant pairing force~\cite{Sano1963}. The more evolved Bogoliubov approach has later been applied to finite nuclei confirming the existence of such relations between $T_c$ and $\Delta(0)$. For the simplified degenerate model, the relation was found to be $T_c =  0.50\Delta(0)$~\cite{Goodman1981}, while in rare earth transition nuclei the interplay with deformation induces shape transitions in the superfluid phase, leading to the ratio $T_c/\Delta(0)=0.57$ for protons and 0.63 for neutrons~\cite{Goodman1986}. In addition, pairing correlations are expected to play an important role in the decay of compound nuclei formed in heavy ion collisions, as illustrated in the seminal work presented in Ref.~\cite{Egido1993}. More recently, the BCS and HFB approaches have been extended to self-consistent mean field models in order to improve
the description of the pairing transition in spherical nuclei~\cite{Reib1999, Khan2007, Niu2013}, as well as in deformed nuclei where shape transitions have been predicted~\cite{Gambhir2000, Egido2000, Agrawal2001, Martin2003, Gambacurta2013}. In summary, the ratio $T_c/\Delta(0)$ lies in the interval $0.50-0.60$, where the uncertainty originates mainly from the detailed level structure of spherical and deformed nuclei which depends itself on models.

It is worth noticing that in most of the quoted studies, the calculations were performed either on the harmonic oscillator basis, or within the non-relativistic framework. Due to the limitation of the harmonic oscillator basis in giving an appropriate asymptotic behavior of the single particle (s.p.) wave functions, the nucleon densities at large distance converge very slowly with respect to the size of the basis. The situation becomes even more serious in the weakly bound nuclei close to the drip line~\cite{Dobaczewski1984, Stoitsov1998}. Nowadays a realistic framework is to perform the calculations in an appropriate basis that can provide a reasonable description of both the overall and asymptotic behaviors of the density profiles, for instance the Woods-Saxon (WS) basis~\cite{Zhou2003, Schunck2008}. In some applications, the small component of the Dirac spinors were usually neglected in determining the relativistic Hartree-Bogoliubov (RHB) and finite-temperature RHB (FT-RHB) pairing tensor~\cite{Niu2013, Niksic2014}. We therefore present, in this paper, the first fully RHFB calculations at finite temperature (FT-RHFB): both the large and small components of the Dirac spinors contribute to the pairing channel, and the contributions of the Fock terms are naturally included. In addition, the Dirac WS (DWS) basis~\cite{Zhou2003} is employed to better describe weakly bound nuclei.

This paper is organized as follows. The general formalism of the FT-RHFB theory is outlined in Sec.~\ref{formalism}. In Sec.~\ref{results}, we compare the results obtained with different covariant density functional (CDF) models~\cite{Bouyssy1987, Ring1996, Bender2003, Vretenara2005, Meng2006, Long2006} using the finite- and zero-range pairing interactions~\cite{Berger1984, Tajima1993}. Systematic FT-RHFB calculations are performed for several isotopes/isotones. Moreover, the questions of pairing persistence and re-entrance phenomena with increasing temperature are addressed. Finally, the main conclusions are drawn in Sec.~\ref{summary}.

\section{GENERAL FORMALISM AND NUMERICAL DETAILS}\label{formalism}

We briefly recall here the general features of RHFB theory and thermodynamics, then their generalization to the finite temperature case.

\subsection{RHFB framework}

The relativistic Hartree-Fock (RHF) theory~\cite{Bouyssy1987, Long2006, Long2007} is designed to describe bulk and s.p. nuclear properties. The relevant degrees of freedom are the isospin doublet nucleon field $\psi_{n(p)}$, the meson fields which mediate the nuclear interactions (two isoscalar fields $\sigma$ and $\omega$, two isovector fields $\rho$ and $\pi$), and the photon field ($A$) that accounts for the electro-magnetic interaction. A general effective Lagrangian is constructed in terms of various currents or densities: $(\bar{\psi} \tau \Gamma \psi)$ where $\tau \in \{1, \bm{\tau}\}$ ($1$ stands for the isoscalar quantities and the isospin Pauli matrix $\bm{\tau}$ corresponds to the isovector ones), and $\Gamma$ denotes the Dirac matrices $\{1, \gamma_\mu, \gamma_5\gamma_\mu, \sigma_{\mu\nu} \}$. Applying the standard variations of the Lagrangian, one may obtain the field equations for nucleon, meson and photon fields (resp. the Dirac, Klein-Gordon and Proca equations), and the continuity equations leading to the Hamiltonian. With the creation and annihilation operators $(c^\dag, c)$ defined from stationary solutions of the Dirac equation, the effective Hamiltonian $\mathcal{H}$ is formally expressed in the second quantized form as
\begin{eqnarray}
\mathcal{H}=\sum_{ij}c^\dagger_i c_j T_{ij}+\frac{1}{2}\sum_{ijkl;\phi}c^\dagger_i c^\dagger_j c_l c_k V^{\phi}_{ijkl},
\end{eqnarray}
where $T_{ij}$ represents the Dirac kinetic energy, and the two-body terms $V_{ijkl}$ correspond to different types of meson- (or photon-) nucleon couplings denoted by $\phi$. In the mean-field approximation, the energy functional $\mathcal{E}$ is obtained by taking the expectation value of the Hamiltonian $\mathcal{H}$ on a Slater determinant, where both the Hartree and Fock terms are considered. The RHFB theory~\cite{Kucharek1991, Long2010, Ebran2011} is deduced by incorporating the Bogoliubov transformation
\begin{eqnarray}
\begin{pmatrix} \alpha \\[0.4em] \alpha^\dagger \end{pmatrix} = \begin{pmatrix} \psi_U^\ast & \psi_V^\ast\\[0.4em]  \psi_V & \psi_U \end{pmatrix} \begin{pmatrix} c \\[0.4em] c^\dagger \end{pmatrix}
\end{eqnarray}
with the RHF model. It provides a unified and self-consistent description of both particle-hole ($ph$) and particle-particle ($pp$) channels, at the mean field level.

\subsection{Thermodynamics and statistical mechanics}

The thermodynamical properties of a system are calculated here in the canonical ensemble.
For a statistical $N$-body system at finite temperature $T$, the equilibrium state is obtained from the variational principle applied to the grand canonical potential $\Omega$~\cite{Goodman1981, Bonche1984, Egido1993},
\begin{eqnarray}
\Omega (T, \lambda) = F - \lambda N = E - TS -\lambda N,
\end{eqnarray}
where $F$ is the free energy, $S$ the entropy, $E$ the total energy, and $\lambda$ the associated Lagrange multiplier.
Namely, the variation
\begin{eqnarray}
\delta\Omega=0
\end{eqnarray}
defines the density operator $\mathcal{D}$ with trace equal to 1, and the grand partition function $Z$,
respectively read as:
\begin{subequations}
\begin{eqnarray}
\mathcal{D} &=& Z^{-1}\text{exp}\{{-\beta(\mathcal{H}-\lambda \mathcal{N})}\},\\
Z &=& \text{Tr}[\text{exp}\{{-\beta(\mathcal{H}-\lambda \mathcal{N})}\}],
\end{eqnarray}
\end{subequations}
where $\mathcal{N}$ is the particle number operator, and $\beta=T^{-1}$. As it is conventional, the temperature $T$ is given in energy units. For arbitrary operator $\mathcal O$, the thermal average over the excited states populated at finite temperature is defined as:
\begin{eqnarray}
\langle\mathcal{O}\rangle = \text{Tr}(\mathcal{D} \mathcal{O}),
\end{eqnarray}
where the trace is taken over all possible excited states.
At finite temperature, the total energy, entropy and number of particles are therefore expressed as
\begin{subequations}
\begin{eqnarray}
E &=& \langle\mathcal{H}\rangle = \text{Tr}(\mathcal{D}\mathcal{H}),\\
S &=& -\langle \text{ln}\mathcal{D}\rangle
= -\text{Tr}(\mathcal{D}\text{ln}\mathcal{D}),\\
N &=& \langle\mathcal{N}\rangle = \text{Tr}(\mathcal{D}\mathcal{N}).
\end{eqnarray}
\end{subequations}

\subsection{Extension to finite temperature}

The FT-RHFB theory is a straightforward generalization of the RHFB theory that readily incorporates a statistical ensemble of excited states.
At the finite-temperature mean field level, the density operator is approximated by~\cite{Goodman1981}:
\begin{eqnarray}
\mathcal{D} = \prod_\alpha \big[f_\alpha \mathcal{N}_\alpha + (1-f_\alpha) (1-\mathcal{N}_\alpha)\big],
\end{eqnarray}
where $f_\alpha$ is the Fermi-Dirac distribution
\begin{eqnarray}
f_\alpha = \langle \mathcal{N}_\alpha \rangle = \frac{1}{1+ e^{\beta E_\alpha}}.
\end{eqnarray}
Notice that, for $T = 0$ we have $f_\alpha = 0$ for all states $\alpha$.
The quasiparticle energy $E_\alpha$ is obtained as the solution of the FT-RHFB equations, see Eq.~(\ref{eq:ftrhfb}).

In spherically symmetric systems the Dirac-Bogoliubov spinor can be written as
\begin{subequations}
\begin{align}\label{Dirac spinor}
\psi_{U_\alpha}(\pmb{r}) &= \frac{1}{r} \begin{pmatrix} iG_{U_a}(r) \\[0.4em] F_{U_a}(r)\pmb{\sigma}\cdot\hat{\pmb{r}} \end{pmatrix}\mathscr{Y}^l_{jm}(\hat{\pmb{r}})\chi_{\frac{1}{2}}(\tau),\\
\psi_{V_\alpha}(\pmb{r}) &= \frac{1}{r} \begin{pmatrix} iG_{V_a}(r) \\[0.4em] F_{V_a}(r)\pmb{\sigma}\cdot\hat{\pmb{r}} \end{pmatrix} \mathscr{Y}^l_{jm}(\hat{\pmb{r}})\chi_{\frac{1}{2}}(\tau),
\end{align}
\end{subequations}
where $G_U$ and $F_U$ correspond to the radial parts of the upper and lower components, respectively,
$\mathscr{Y}^l_{jm}(\hat{\pmb{r}})$ is the spinor spherical harmonics, and $\chi_{\frac{1}{2}}(\tau)$ is the isospinor.
Here, the subindex $\alpha = (a,m) = (n, l, j, m)$ contains the quantum numbers $n$ (number
of nodes of the upper component $G_{U(V)}$), $l$ (orbital angular momentum), and $j, m$ (total angular
momentum and its projection to the $z$ axis).

The minimization of $\Omega$ with respect to $\mathcal{D}$ leads to the FT-RHFB equations,
\begin{align}\label{eq:ftrhfb}
&\int d\pmb r^\prime \begin{pmatrix} h(\pmb{r}, \pmb{r}^\prime) & \Delta (\pmb{r}, \pmb{r}^\prime)\\[0.4em] -\Delta(\pmb{r}, \pmb{r}^\prime) & h(\pmb{r}, \pmb{r}^\prime)\end{pmatrix}\begin{pmatrix}\psi_U(\pmb{r}^\prime) \\[0.4em] \psi_V(\pmb{r}^\prime)\end{pmatrix} \nonumber\\
&\hspace{8em}= \begin{pmatrix}\lambda+E &0\\[0.4em] 0 &\lambda-E\end{pmatrix} \begin{pmatrix}\psi_U(\pmb{r}) \\[0.4em] \psi_V(\pmb{r})\end{pmatrix}
\end{align}
which are formally identical to the RHFB equations~\cite{Kucharek1991, Long2010, Ebran2011}. The Dirac Hamiltonian $h(\pmb{r}, \pmb{r}^\prime)$ contains the kinetic energy $h^K$, the direct local potential $h^D$, and exchange non-local potential $h^E$,
\begin{subequations}
\begin{eqnarray}
h^K(\pmb{r},\pmb{r}^\prime) &=& \big[\pmb{\alpha}\cdot\pmb{p}+\beta M\big]\delta(\pmb{r}-\pmb{r}^\prime),\\
h^D(\pmb{r},\pmb{r}^\prime) &=& \big[\Sigma_T(\pmb{r})\gamma_5+\Sigma_0(\pmb{r})+\beta\Sigma_S(\pmb{r})\big]\delta(\pmb{r}-\pmb{r}^\prime),\\
h^E(\pmb{r},\pmb{r}^\prime) &=& \begin{pmatrix}Y_G(\pmb{r},\pmb{r}^\prime)&Y_F(\pmb{r},\pmb{r}^\prime)\\[0.4em]
                                   X_G(\pmb{r},\pmb{r}^\prime)&X_F(\pmb{r},\pmb{r}^\prime)\end{pmatrix}.
\end{eqnarray}
\end{subequations}
In the above expressions, the local self-energies $\Sigma_S$, $\Sigma_0$, and $\Sigma_T$ contain the contributions from the Hartree (direct) terms and the rearrangement terms,
which depend directly on various local quasiparticle densities. They are the vector, scalar and tensor densities, respectively,
\begin{subequations}
\begin{eqnarray}
\rho_v (r)&=&\frac{1}{4\pi r^2}\sum_a\hat{j}^2_a
\Big\{\big[ G^2_{V_a}(r) + F^2_{V_a}(r)\big](1-f_a) \nonumber\\
&& + \big[G^2_{U_a}(r) + F^2_{U_a}(r)\big] f_a\Big\},\label{eq:13a}\\
\rho_s (r)&=&\frac{1}{4\pi r^2}\sum_a\hat{j}^2_a
\Big\{\big[ G^2_{V_a}(r) - F^2_{V_a}(r)\big](1-f_a) \nonumber\\
&& + \big[G^2_{U_a}(r) - F^2_{U_a}(r)\big] f_a\Big\},\\
\rho_t (r)&=&\frac{1}{4\pi r^2}\sum_a\hat{j}^2_a
\Big\{2G_{V_a}(r)F_{V_a}(r)(1-f_a) \nonumber\\
&& + 2G_{U_a}(r)F_{U_a}(r) f_a\Big\}.\label{eq:13c}
\end{eqnarray}
\end{subequations}
Notice that, at the limit $T=0$, where $f_a=0$ Eqs.~(\ref{eq:13a})-(\ref{eq:13c}) give back the usual expressions~\cite{Long2010}.
The nonlocal self-energies $X_{G(F)}$ and $Y_{G(F)}$ come from the Fock (exchange) terms,
\begin{subequations}\label{eq:nolocal}
\begin{eqnarray}
X^{(\phi)}_{G_a}(r,r^\prime)&=&\sum_b\mathscr{T}^\phi_{ab}\hat{j}^2_b
\Big[(g_\phi F_{V_b})_r\mathscr{R}^{X_G}_{ab}(m_\phi;r,r^\prime)(g_\phi G_{V_b})_{r^\prime}(1-f_b) \nonumber\\
&& + (g_\phi F_{U_b})_r\mathscr{R}^{X_G}_{ab}(m_\phi;r,r^\prime)(g_\phi G_{U_b})_{r^\prime}f_b\Big],\\
X^{(\phi)}_{F_a}(r,r^\prime)&=&\sum_b\mathscr{T}^\phi_{ab}\hat{j}^2_b
\Big[(g_\phi F_{V_b})_r\mathscr{R}^{X_F}_{ab}(m_\phi;r,r^\prime)(g_\phi F_{V_b})_{r^\prime}(1-f_b) \nonumber\\
&& + (g_\phi F_{U_b})_r\mathscr{R}^{X_F}_{ab}(m_\phi;r,r^\prime)(g_\phi F_{U_b})_{r^\prime}f_b\Big],\\
Y^{(\phi)}_{G_a}(r,r^\prime)&=&\sum_b\mathscr{T}^\phi_{ab}\hat{j}^2_b
\Big[(g_\phi G_{V_b})_r\mathscr{R}^{Y_G}_{ab}(m_\phi;r,r^\prime)(g_\phi G_{V_b})_{r^\prime}(1-f_b) \nonumber\\
&& + (g_\phi G_{U_b})_r\mathscr{R}^{Y_G}_{ab}(m_\phi;r,r^\prime)(g_\phi G_{U_b})_{r^\prime}f_b\Big],\\
Y^{(\phi)}_{F_a}(r,r^\prime)&=&\sum_b\mathscr{T}^\phi_{ab}\hat{j}^2_b
\Big[(g_\phi G_{V_b})_r\mathscr{R}^{Y_F}_{ab}(m_\phi;r,r^\prime)(g_\phi F_{V_b})_{r^\prime}(1-f_b) \nonumber\\
&& + (g_\phi G_{U_b})_r\mathscr{R}^{Y_F}_{ab}(m_\phi;r,r^\prime)(g_\phi F_{U_b})_{r^\prime}f_b\Big].
\end{eqnarray}
\end{subequations}
In these expressions, $\mathscr{T}^\phi_{ab}$ denotes the isospin factors:
$\delta_{ab}$ for isoscalar channels and $2-\delta_{ab}$ for isovector channels,
$\hat{j}^2_b = 2j_b + 1$ is the degeneracy number of the corresponding energy level, $g_\phi$ represents the coupling constants, and $\mathscr{R}_{ab}$ denote the multipole expansions of meson propagators.
Eq.~(\ref{eq:nolocal}) is a generalization at finite temperature of the expressions given in Ref.~\cite{Long2010}.

Next, we consider the pairing field $\Delta(r, r^\prime)$ of Eq.~(\ref{eq:ftrhfb}),
\begin{eqnarray}
\Delta_a(r, r^\prime) = -\frac{1}{2}\sum_b V^{pp}_{ab}(r,r^\prime)\kappa_b(r,r^\prime)
\end{eqnarray}
with the pairing interaction $V^{pp}$ and the pairing tensor $\kappa$.
If we take a finite-range pairing force, the pairing tensor $\kappa$ will read as
\begin{eqnarray}
\kappa_a (r, r^\prime)
&=&\hat{j}_a^2\Big\{[G_{V_a}(r)G_{U_a}(r^\prime)+F_{V_a}(r)F_{U_a}(r^\prime)] \\
& & +[G_{U_a}(r)G_{V_a}(r^\prime)+F_{U_a}(r)F_{V_a}(r^\prime)]\Big\}(1-2f_a).\nonumber
\end{eqnarray}
Notice that the temperature dependence of the solution $(E_\alpha;\psi_{U_\alpha}, \psi_{V_\alpha})$ of the FT-RHFB Eq.~(\ref{eq:ftrhfb}) comes implicitly through the quasiparticle densities $\rho_{v, s, t}$, nonlocal potentials $X(Y)$ and pairing tensor $\kappa$.

For practical evaluation of the pairing correlations, an average pairing gap is introduced and defined as the ratio
of the pairing energy over the pairing tensor,
\begin{eqnarray}\label{eq:deltauv}
\Delta = \frac{\text{Tr}(\Delta\kappa)}{\text{Tr}\kappa}.
\end{eqnarray}
This quantity is calculated for neutrons ($\Delta_n$) and protons ($\Delta_p$) separately
and it is discussed in detail in the next section.

The total FT-RHFB energy $E$ of the system is calculated with the microscopic two-body center-of-mass
correction~\cite{Bender2000}.
The entropy $S$ of the system can be evaluated from
\begin{eqnarray}
S(T) = - \sum_\alpha \Big[f_\alpha \ln f_\alpha
+ (1- f_\alpha)\ln (1- f_\alpha)\Big],
\end{eqnarray}
and the specific heat is defined by
\begin{eqnarray}
C_v(T) = T\frac{\partial S(T)}{\partial T}\Big\vert_{N}.
\end{eqnarray}
They correspond to the first and second derivative of the free energy $F$, respectively.

The integro-differential FT-RHFB Eq.~(\ref{eq:ftrhfb}) is solved by using a DWS basis~\cite{Zhou2003} with a radial cut off $R=26$ fm. The numbers of positive and negative energy
states in the basis expansion for each s.p. angular momentum $(l,j)$ are chosen to be 36 and 12, respectively.
We have verified that this truncation scheme provides sufficient numerical accuracy for the description of weakly
bound nuclei (Pb isotopes).

\section{RESULTS AND DISCUSSION}\label{results}

\begin{table*}[tb]
\caption{
Bulk properties of symmetric nuclear matter at the saturation point: density $\rho_0$ (fm$^{-3}$), binding energy $E_{B}/A$ (MeV), compression
modulus $K$ (MeV), symmetry energy $J$ (MeV) and non-relativistic effective masses $M^\ast_{NR}$ (M) predicted by selected RHF and RH models. The non-relativistic effective masses in neutron matter are also listed.}
\setlength{\tabcolsep}{8pt}
\label{tab:NMP}
\begin{tabular}{llccccccccc}
\hline\hline
\multirow{2}*{Model} & \multirow{2}*{Interaction} & \multirow{2}*{Ref.} & \multicolumn{5}{c}{symmetric matter} & & \multicolumn{2}{c}{neutron matter} \\
\cline{4-8}\cline{10-11} & & & $\rho_0$ & $E_{B}/A$ & $K$ & $J$ & $M^\ast_{NR}$ & & $M^\ast_{NR}(\nu)$ & $M^\ast_{NR}(\pi)$  \\
\hline
    & PKA1  & \cite{Long2007} & 0.160 & $-15.83$ &  229.96 &  36.02          &  0.68         & & 0.68 & 0.70  \\
RHF & PKO1  & \cite{Long2006} & 0.152 & $-16.00$ &  250.28 &  \textbf{34.37} &  0.75         & & 0.73 & 0.76  \\
    & PKO2  & \cite{Long2008} & 0.151 & $-16.03$ &  249.53 &  \textbf{32.49} & \textbf{0.76} & & 0.75 & 0.77  \\
\\
    & DD-ME2  & \cite{Lalazissis2005} & 0.152 & $-16.14$ &  250.97 &  32.31 & \textbf{0.65} & & 0.64 & 0.70  \\
RH  & PK1r    & \cite{Long2004} & 0.148 & $-16.27$ &  \textbf{283.68} &  37.83 & 0.68 & & 0.64 & 0.72        \\
    & NL3$^*$ & \cite{Lalazissis2009} & 0.150 & $-16.30$ &  \textbf{258.56} &  38.70 & 0.67 & & 0.63 & 0.72  \\
\hline\hline
\end{tabular}
\end{table*}

In this section, we compare the predictions based on several models which are determined by the model
Lagrangian as well as the pairing interaction.
The latter interaction is either a finite-range Gogny D1S interaction~\cite{Berger1984} or a density-dependent
contact interaction (DDCI) of the form~\cite{Tajima1993}:
\begin{equation}
V(r, r^\prime)=V_0\frac{1}{2}(1-P_\sigma)\left( 1-\frac{\rho(r)}{\rho_0} \right) \delta(r-r^\prime).
\end{equation}
Notice that the DDCI requires a regularisation scheme.
We have considered in this study a simple cut-off scheme, defined to be 100~MeV in quasiparticle space, and
the strength $V_0$ is adjusted to reproduce the same average pairing gap as that obtained with the Gogny D1S interaction.
These values of $V_0$ will be given with the results hereafter.
The D1S pairing interaction depends slightly on the mass:
A general factor $g$ is therefore introduced for its strength,
as in Refs.~\cite{Wang2013, Agbemava2014}.
We consider 6 different model Lagrangians which are given in Table~\ref{tab:NMP}.
Some of the bulk properties of nuclear matter determined by these Lagrangians are also shown.
They are generally compatible with the expected gross properties of finite nuclei, e.g., the binding energy
$E/A \simeq 16$ MeV, the saturation density
$\rho_0 \simeq 0.15$ fm$^{-3}$, the incompressibility modulus $K \in[220, 280]$ MeV, and
the symmetry energy $J \in[32, 39]$ MeV.

In the following discussion, the persistence and re-entrance of the pairing phenomenon will be
commented and analysed. Let us briefly recall the unifying mechanism which is at play in these various phenomena,
recently discussed in Ref.~\cite{Margueron2012a}.
Due to thermal excitations, s.p. states above the Fermi energy can be slightly populated while states below the Fermi
energy can be partially depleted.
This occurs if the involved new states are not too far in energy from the last occupied state, but it should also
be not too close, otherwise, these states would already participate to pairing correlations at zero temperature.
The typical shell gap should be around 2~MeV.
The participation of these states at finite temperature gives rise either to the persistence of pairing correlations
slightly above the usual critical temperature for nuclei which are already superfluid at zero temperature, or
to pairing re-entrance at finite temperature for nuclei which have weak or no pairing at zero temperature.
The best nuclei, in which such a phenomenon is expected, are those close to the drip line, as well as those located at a subshell closure as shown in this section.

In the following subsections, we evaluate the influence of the model on the pairing properties in hot finite nuclei,
taking $^{124}$Sn as an example for testing pairing correlations.
The ratio of the critical temperature over the average pairing gap at zero temperature $T_c/\Delta_n(0)$ is also
studied as a function of the model, varying either the Lagrangian or the type of pairing interaction.
Finally, a more systematic study of the critical temperature is performed on a set of semi-magic nuclei.

\subsection{Study of the ratio $T_c/\Delta_n(0)$}
\begin{figure}[tb]
\centering
%\ifpdf
%\includegraphics[width = 0.48\textwidth]{Mnr.pdf}
%\else
\includegraphics[width = 0.48\textwidth]{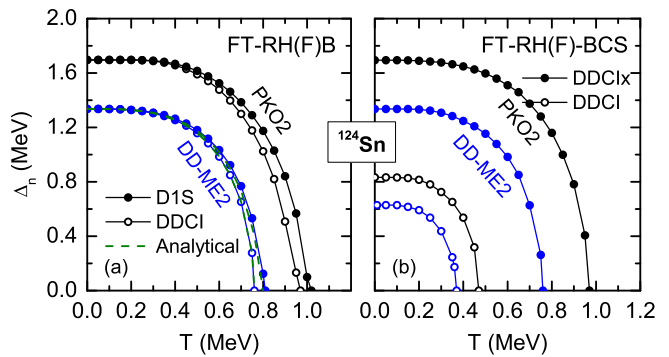}
%\fi
\caption{(Color online)
The neutron pairing gaps in $^{124}$Sn as a function of temperature,
calculated with PKO2 (black) and DD-ME2 (blue) corresponding to different $M^\ast_{NR}$.
In the pairing channel, we compare finite-range D1S (filled circles) and
contact DDCI (empty circles) forces.
The analytical results (dashed lines) are also shown.
Notice that here, we took the factor $g=1$ for the Gogny D1S force, and for the DDCI pairing, we have
taken the following values for $V_0$ (in MeV.fm$^{-3}$): 335 (PKO2), and 342 (DD-ME2) for the RH(F)B and BCS
calculations.
For the DDCIx force used in BCS, we took for $V_0$ (in MeV.fm$^{-3}$): 526 (PKO2) and 539 (DD-ME2).
}
\label{fig:MNR}
\end{figure}

Since pairing correlations are active only around the Fermi level, the ratio $T_c/\Delta_n(0)$ is expected to be modified by the effective mass which influences the s.p. level spacing to a large extent.
Notice that here, the effective mass corresponds to the non-relativistic one $M^\ast_{NR}$
instead of the quantity $M^\ast_S = M+\Sigma_S$ that is named as the Dirac mass~\cite{Jaminon1989, Long2006}.
It can indeed be shown that, in the weak coupling limit of the BCS approximation, the average pairing gap
at the Fermi surface $\Delta_F$ can be expressed as~\cite{Baldo1999}
\begin{equation}
\Delta_F \approx 2 \epsilon_F \exp\big[2/(N_F v_{\text{pair}})\big],
\label{eq:deltaf}
\end{equation}
where $\epsilon_F$ is the Fermi energy, $N_F=m^\ast_F k_F/\pi^2$ is the average density of state in uniform matter at the Fermi energy, and $k_F$ denotes the Fermi momentum, and $v_{\text{pair}}$ is a constant pairing interaction. It is clear from Eq.~(\ref{eq:deltaf}) that the pairing gap $\Delta_F$ is quite sensitive to the effective mass at the Fermi energy $\epsilon_F$. Eq.~(\ref{eq:deltaf}) is obtained in infinite matter and it provides only a qualitative understanding of the relation between the pairing force strength and the effective mass. In the following, we present a quantitatively precise analysis of the correlation between the critical temperature and the non-relativistic effective mass in finite nuclei.

Figure~\ref{fig:MNR}(a) displays the evolution of the neutron pairing gap as a function of temperature for $^{124}$Sn, a good candidate for studying pairing correlations. We compare two Lagrangians, PKO2 and DD-ME2 (see Table~\ref{tab:NMP}) both with two kinds of pairing interactions: the finite-range Gogny D1S force~\cite{Berger1984} and the contact force DDCI~\cite{Tajima1993}. The two Lagrangians PKO2 and DD-ME2 mostly differ by their non-relativistic effective mass $M^\ast_{NR}$ (see Table~\ref{tab:NMP}), and it is observed that the average pairing gap at zero temperature $\Delta_n$ scales with $M^\ast_{NR}$, as expected from the weak coupling expression~(\ref{eq:deltaf}). Comparing the different types of pairing interaction (finite- or zero-range) for the same Lagrangian, it is observed in Fig.~\ref{fig:MNR} that the vanishing of pairing correlations at finite temperature slightly depends on the type of pairing force, namely, with the zero-range pairing interaction, the critical temperature $T_c$ is slightly lower than with the finite-range interaction. From Fig.~\ref{fig:MNR}, the ratio $T_c/\Delta_n(0)$ is obtained as 0.60 for D1S, and 0.57 for DDCI. In Ref.~\cite{Dobaczewski1996}, it was shown that the dependence of the pairing gap on the state around the Fermi energy is qualitatively different for contact and finite-range interactions. In addition, with a finite-range interaction, the pairing gap and the pairing tensor have non-local components which cannot be simply absorbed in the DDCI. The slight increase of the critical temperature with the D1S interaction is therefore an effect of the finite-range nature of the interaction, dispersing the pairing effects among more s.p. states. For comparison, and taking $T_c=0.60\Delta_n(0)$ (finite-range pairing force) and $T_c=0.57\Delta_n(0)$ (zero-range pairing forces), the analytical relation~\cite{Goriely1996},
\begin{equation}\label{eq:analy}
\Delta_n(T)=\Delta_n(0) \Bigg[1-\Big(\frac{T}{T_c}\Big)^m \Bigg]^{1/2}\Theta(T-T_c)
\end{equation}
where $m=3.32$, is plotted for the DD-ME2 model [dashed lines in Fig.~\ref{fig:MNR}(a)]. There is almost no difference between the analytical model, i.e., Eq.~(\ref{eq:analy}), and the numerical calculations for stable nuclei like $^{124}$Sn.

Figure~\ref{fig:MNR}(b) shows the neutron pairing gap calculated with the RHF theory plus BCS pairing at finite temperature (FT-RHF-BCS), using the same DDCI interaction as in the
FT-RHFB calculations shown in Fig.~\ref{fig:MNR}(a) and modified one (DDCIx) with enhanced pairing strength parameter $V_0$ (see caption for more details). It is found that, with the same DDCI pairing interaction, the neutron pairing gaps determined by the BCS method are reduced to about half of Bogoliubov results. Such distinct difference is due to the fact that the off-diagonal couplings, which account for about half of the pairing correlations, are absent in the BCS pairing. As a result, the strength parameter $V_0$ in BCS calculations is usually larger than in HFB, see for instance Ref.~\cite{Bertsch2009}. We have therefore readjusted $V_0$ in the RHF-BCS calculation at zero temperature to obtain the same pairing gap as the RHFB prediction, leading to the DDCIx interaction in Fig.~\ref{fig:MNR}(b). Applying such a simple modification of the parameter $V_0$ in $^{124}$Sn, the temperature dependence of the pairing gap predicted by the FT-RHFB and FT-RHF-BCS frameworks are almost undistinguishable. However, this is not always true and as we will see that the above simple renormalization of the pairing strength will not work towards the drip line where the coupling to continuum states becomes more and more important.

\begin{figure}[b]
\centering
%\ifpdf
%\includegraphics[width = 0.48\textwidth]{SETs.pdf}
%\else
\includegraphics[width = 0.48\textwidth]{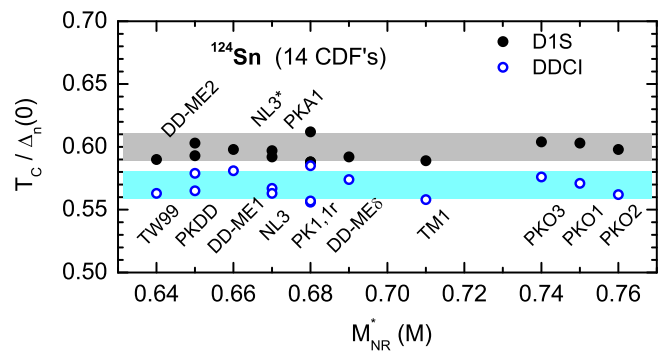}
%\fi
\caption{(Color online)
The ratios $T_c / \Delta_n (0)$ in $^{124}$Sn as a function of the non-relativistic
effective mass $M^\ast_{NR}$, calculated using the FT-RH(F)B theory with 14 parameter sets.
In the pairing channel the finite-range D1S and the contact force DDCI are employed.
}
\label{fig:SETs}
\end{figure}

The critical temperatures is also expected to depend on the non-relativistic effective mass, see for instance Refs.~\cite{Reib1999, Gambhir2000, Yuksel2014}.
The dependence of the ratio $T_c/\Delta_n(0)$ on the effective mass is, however, not very well known. In Fig.~\ref{fig:SETs}, we plot the ratio $T_c/\Delta_n(0)$ as a function of $M^\ast_{NR}$ in $^{124}$Sn, and we compare the predictions of two pairing interactions (finite versus zero range). The effective mass $M_{NR}^*$ is obtained with 14 CDFs (the 6 CDFs given in Table~\ref{tab:NMP} completed with 8 other CDFs: PKO3~\cite{Long2008}, PKDD~\cite{Long2004}, DD-ME1~\cite{Niksic2002}, DD-ME$\delta$~\cite{Roca2011}, TW99~\cite{Typel1999}, PK1~\cite{Long2004}, NL3~\cite{Lalazissis1997} and TM1~\cite{Sugahara1994}). It is found that the ratio $T_c/\Delta_n(0)$ does not depend much on $M^\ast_{NR}$. Similar analyses were also carried out to check the relations with the incompressibility modulus $K$ and the symmetry energy $J$, but no evidence of correlation is found. The small fluctuations observed in Fig.~\ref{fig:SETs} are therefore mostly shell effects. In conclusion, the critical temperature in stable nuclei scales very well with the average pairing gap at zero temperature, the shell effects contributing to a dispersion of less than $4\%$.

\begin{figure}[tb]
%\ifpdf
%\includegraphics[width = 0.48\textwidth]{TC.pdf}
%\else
\includegraphics[width = 0.48\textwidth]{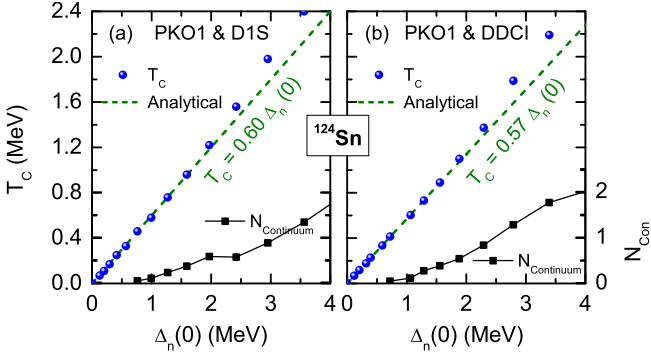}
%\fi
\caption{(Color online)
The critical temperature $T_c$ (left axes, blue circles) and the occupation number of continuum states $N_{con}$ (right axes, black squares) at zero temperature in $^{124}$Sn as a function of neutron pairing gap calculated from PKO1 FT-RHFB with Gogny D1S (a) and DDCI (b) pairing forces. The dashed lines correspond to the relation $T_c= 0.60(0.57)\Delta_n(0)$.
The strength parameter of the pairing interaction is varied to obtain different values of the pairing gap at zero temperature.
}
\label{fig:TC}
\end{figure}

In order to check whether the ratio $T_c/\Delta_n(0)$ is still constant for larger pairing strength,
we have artificially varied the pairing interaction strength and correlated the critical temperature $T_c$
with the neutron pairing gap $\Delta_n(0)$ at zero temperature. Figures~\ref{fig:TC}(a) and ~\ref{fig:TC}(b)
respectively show the results calculated with the pairing interactions D1S and DDCI using the PKO1 Lagrangian,
where the average pairing gap $\Delta_n(0)$ goes from 100~keV up to about 3.5~MeV.
At low pairing gap, the ratios $T_c/\Delta_n(0)$ are consistent with the analytic ones (dashed lines) as expected,
until small deviations emerge beyond $\Delta_n(0)\sim2.5$~MeV,
where the continuum contributions become sizable (filled squares). These results are consistent with previous findings~\cite{Niu2013} based on a separable version of the Gogny D1S pairing force~\cite{Tian2009} and RH Lagrangian with PC-PK1 point coupling~\cite{Zhao2010}. A simple calculation in infinite matter with a contact pairing interaction gives a correction to the ratio~\cite{Mahan2000}
\begin{equation}
\frac{T_c}{\Delta(0)}\approx 0.57\Big[1-\frac{1}{4\omega_D^2}\Delta(0)^2\Big],
\label{eq:tcunif}
\end{equation}
where $\omega_D$ represents the pairing window. According to Eq.~(\ref{eq:tcunif}), the next-to-leading order correction has a negative sign, at variance with our results in $^{124}$Sn, see Fig.~\ref{fig:TC}. Notice that for the DDCI pairing interaction, the pairing window is about 100~MeV. The correction to the linear approximation of Eq.~(\ref{eq:tcunif}) is therefore very small: for the maximal pairing gap considered in this work ($\Delta(0)\approx 4.0$~ MeV), the correction represents no more than a few percent of the linear leading term.

\begin{figure}[tb]
\centering
%\ifpdf
%\includegraphics[width = 0.30\textwidth]{Cont.pdf}
%\else
\includegraphics[width = 0.30\textwidth]{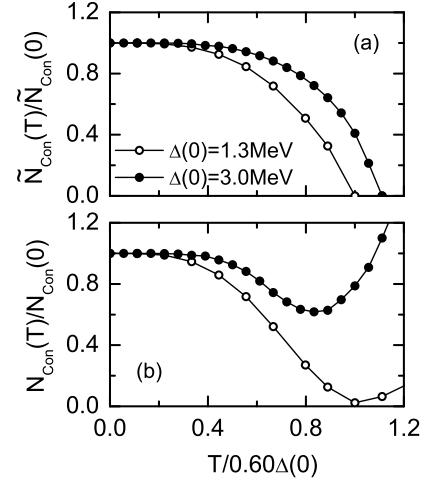}
%\fi
\caption{
Contributions, in $^{124}$Sn, of the continuum states to the pairing number $\tilde{N}_{con}$ (a) and to the neutron number $N_{con}$ (b), normalized to their value at zero temperature, as a function of the temperature $T/0.6\Delta(0)$.
The results correspond to PKO1 Lagrangian and D1S pairing interaction.
}
\label{fig:Cont}
\end{figure}

In Fig.~\ref{fig:TC}, the increase of the critical temperature for $\Delta_n(0)>2.0$~MeV reveals an enhancement of the thermal pairing correlations, as well as an the important role played by the continuum states. We remind that the next-to-the-leading order correction appearing in the simple expression in uniform matter~(\ref{eq:tcunif}) is negative. However Fig.~\ref{fig:TC} shows a continuous enhanced continuum effect with respect to the pairing gap $\Delta_n(0)$.
To better stress the increasing contribution of the continuum states and their role in the pairing correlations, we
have introduced the following two quantities: the cumulative occupation number of the neutron continuum states $N_{con}$,
\begin{eqnarray}
N_{con} = \sum_{a, \epsilon_a \geq 0} \int 4\pi r^2 \rho_a(r) dr,
\end{eqnarray}
and the cumulative pairing-occupation number of the neutron continuum states $\tilde{N}_{con}$,
\begin{eqnarray}
\tilde{N}_{con} = \sum_{a, \epsilon_a \geq 0} \int 4\pi r^2 \kappa_a(r) dr.
\end{eqnarray}
Here the continuum states $a$ are determined in the $T=0$ canonical basis, for simplicity~\cite{Bonche1984, Gambhir2000}, with s.p. energy $\epsilon_a$ above the continuum threshold.
The increasing role of the continuum states at finite temperature is illustrated in Fig.~\ref{fig:Cont} where the contributions to the pairing number $\tilde{N}_{con}$ [plot (a)] and to the neutron number $N_{con}$ [plot (b)] from the continuum states in $^{124}$Sn, normalized to their values at zero temperature, are shown. Results with a weak pairing ($\Delta_n(0)=1.3$~MeV) and with a stronger pairing ($\Delta_n(0)=3.0$~MeV) are compared.
Even if in the latter case the pairing is slightly larger than the expected value in finite nuclei, its inclusion in our analysis helps to understand the role of the continuum states.
For the weak pairing case, the continuum effects are very small, see Fig.~\ref{fig:TC},
and both $N_{con}$ and $\tilde{N}_{con}$ drop to zero at the expected value $T_c=0.6\Delta(0)$, see Fig.~\ref{fig:Cont}.
For the strong pairing case, a clear correlation is observed in Fig.~\ref{fig:Cont} between the increase of the occupation of the continuum states at finite temperature $N_{con}$ [see plot (b)] and the persistence of the pairing numbers $\tilde{N}_{con}$ [see plot (a)].
The persistence of pairing correlations below the critical temperature modifies also the critical temperature, and it is observed in Fig.~\ref{fig:Cont}(a) a larger value of the quantity $T/0.6\Delta(0)$ where pairing correlations in the continuum space drops to zero in the case of strong pairing compared to the weak one.
Coming back to Fig.~\ref{fig:TC}, we now understand better the correlation between the slight deviations of $T_c$ from the analytical behaviors and enhanced continuum effects for $\Delta(0)\gtrsim2.5$~MeV.
Since the presence of resonance states in the continuum is a typical feature of finite systems, the increase observed in Fig.~\ref{fig:TC}, which is different from the prediction of Eq.~(\ref{eq:tcunif}), is then expected only in finite systems. Anticipating the results shown in the next figures, a similar enhancement of pairing correlations in other nuclei will also be observed, revealing here also the role of the resonant states.

\subsection{Evolution of the critical temperature in isotopic and isotonic chains}

\begin{figure}[tb]
\centering
%\ifpdf
%\includegraphics[width = 0.48\textwidth]{GAPa1.pdf}
%\else
\includegraphics[width = 0.48\textwidth]{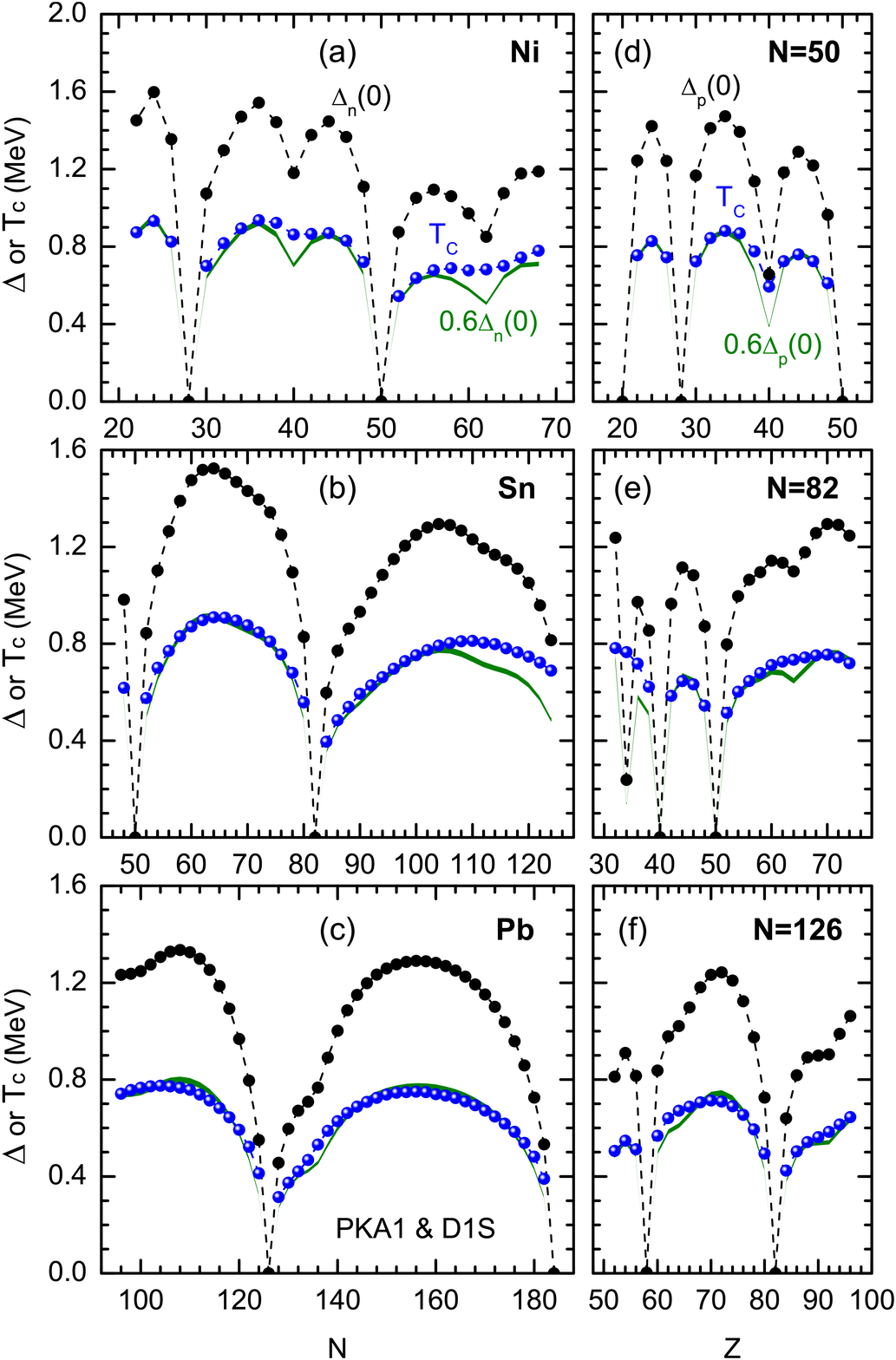}
%\fi
\caption{ (Color online)
Comparison of $\Delta(0)$ (black circles), $0.60\Delta(0)$ (green curves) and $T_c$ (blue circles) in the even-even Ni, Sn, Pb isotopes (left panels) and $N = 50$, 82, 126 isotones (right panels), calculated in FT-RHFB with PKA1 and the Gogny pairing interaction D1S.
}
\label{fig:GAPa1}
\end{figure}

In this subsection, we perform a systematic analysis of the evolution of the pairing gaps and critical
temperature along isotopic and isotonic chains of semi-magic nuclei.
Through this extensive analysis, we have access to various s.p. configurations going from stable nuclei towards
weakly bound drip line nuclei, and we probe the pairing correlations inside various major shells.
We consider three models, PKA1 and PKO1 (RHF) and DD-ME2 (RH) which have different symmetry energies and
non-relativistic effective masses, see Table~\ref{tab:NMP}.
Anticipating our results, we will see that these models lead to rather different predictions for the pairing gap $\Delta$.

Our results are shown for PKA1 (Fig.~\ref{fig:GAPa1}), PKO1 (Fig.~\ref{fig:GAPo1}) and DD-ME2 (Fig.~\ref{fig:GAPme}) models, with the pairing channel described by the D1S interaction. In Figs.~\ref{fig:GAPa1}-{\ref{fig:GAPme}, we have represented isotopic (Ni, Sn and Pb) and isotonic ($N=50$, 82, and 126) average pairing gap evolution as a function of $N$ (for isotopes) and $Z$ (for isotones). The isotopic and isotonic chains are bounded by the drip lines predicted by each of the considered models and determined by the two-nucleon separation energy. These drip lines are consistent, within a few units uncertainties, with predictions given by other models obtained with Skyrme forces~\cite{Erler2012, Erler2013}, Gogny forces~\cite{Hilaire2007, Delaroche2010} and RH Lagrangians~\cite{Afanasjev2013, Agbemava2014}. We have calculated the average pairing gap at zero temperature (filled black circles) defined from Eq.~(\ref{eq:deltauv}), and compared the calculated critical temperature (filled blue circles) with the approximate relation, i.e., $0.6\Delta(0)$ (green curves). The arch structure of the results shown in Figs.~\ref{fig:GAPa1}-{\ref{fig:GAPme} reflect the presence of magic numbers where pairing correlations completely vanish.

The PKA1 model (Fig.~\ref{fig:GAPa1}) is the most complete RHF version of the CDF theory. It contains the $\rho$-$N$ Lorentz tensor coupling which is known to enhance the spin-orbit splitting~\cite{Long2007, Long2009, Li2014}: in many cases the subshell structure is found to be closer to the experimental data than those predicted by other models without the $\rho$-$N$ Lorentz tensor coupling, such as the RH approaches shown in Fig.~\ref{fig:GAPme}. These subshell structures are clearly visible in Figs.~\ref{fig:GAPa1}-{\ref{fig:GAPme} since they induce a partial quenching of the pairing gap for the associated submagic numbers. Going towards the drip lines, a reduction of the pairing gaps is often observed, revealing the presence of closed-shell nuclei at or near the drip lines. For the neutron drip line, it is the case of Sn and Pb isotopes, and for the proton drip line, it is observed for $N=50$.

\begin{figure}[tb]
\centering
%\ifpdf
%\includegraphics[width = 0.48\textwidth]{GAPo1.pdf}
%\else
\includegraphics[width = 0.48\textwidth]{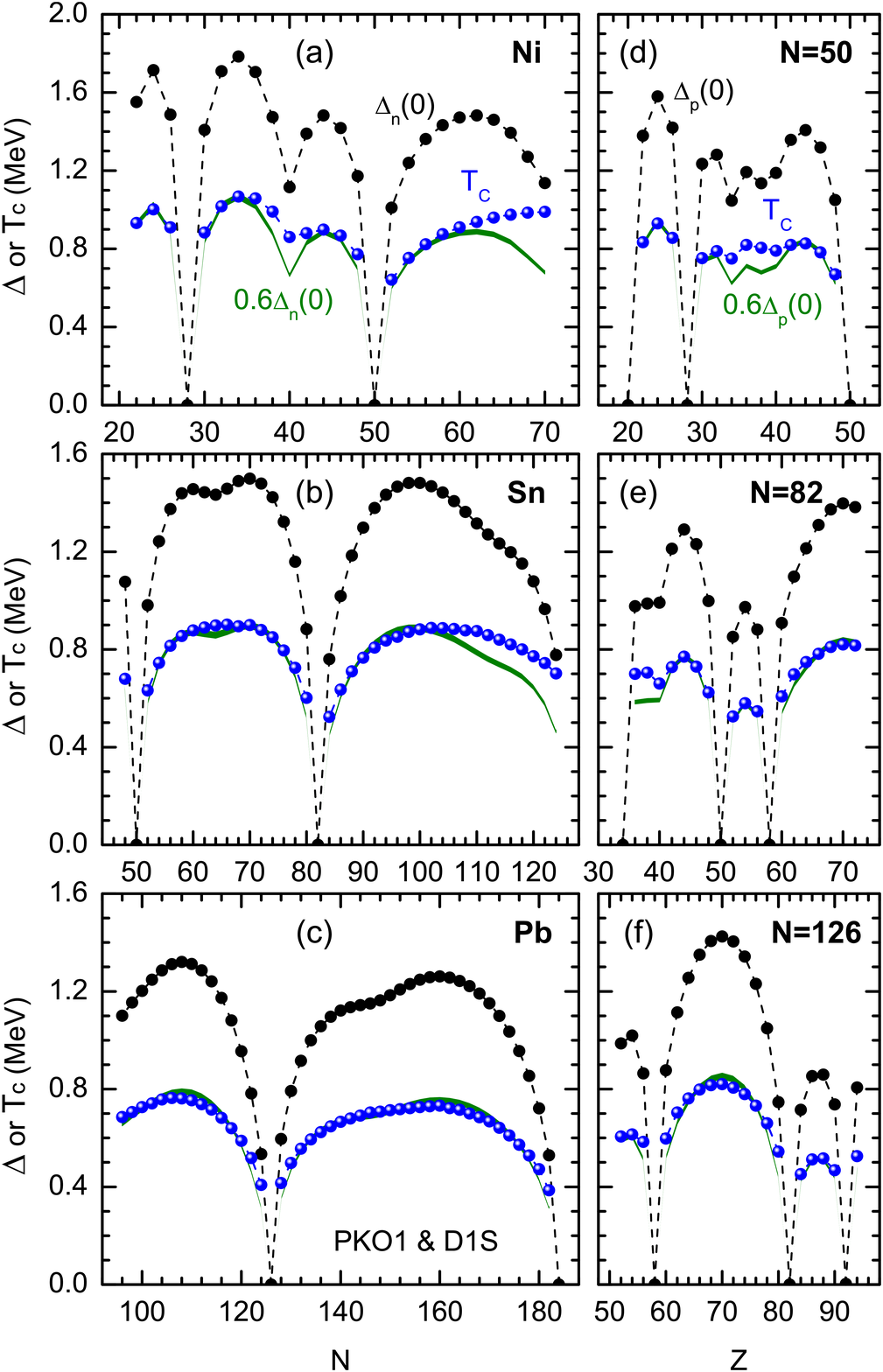}
%\fi
\caption{ (Color online)
The same as in Fig.~\ref{fig:GAPa1}, but calculated in the FT-RHFB theory with the
effective interaction PKO1.
}
\label{fig:GAPo1}
\end{figure}

\begin{figure}[tb]
\centering
%\ifpdf
%\includegraphics[width = 0.48\textwidth]{GAPme.pdf}
%\else
\includegraphics[width = 0.48\textwidth]{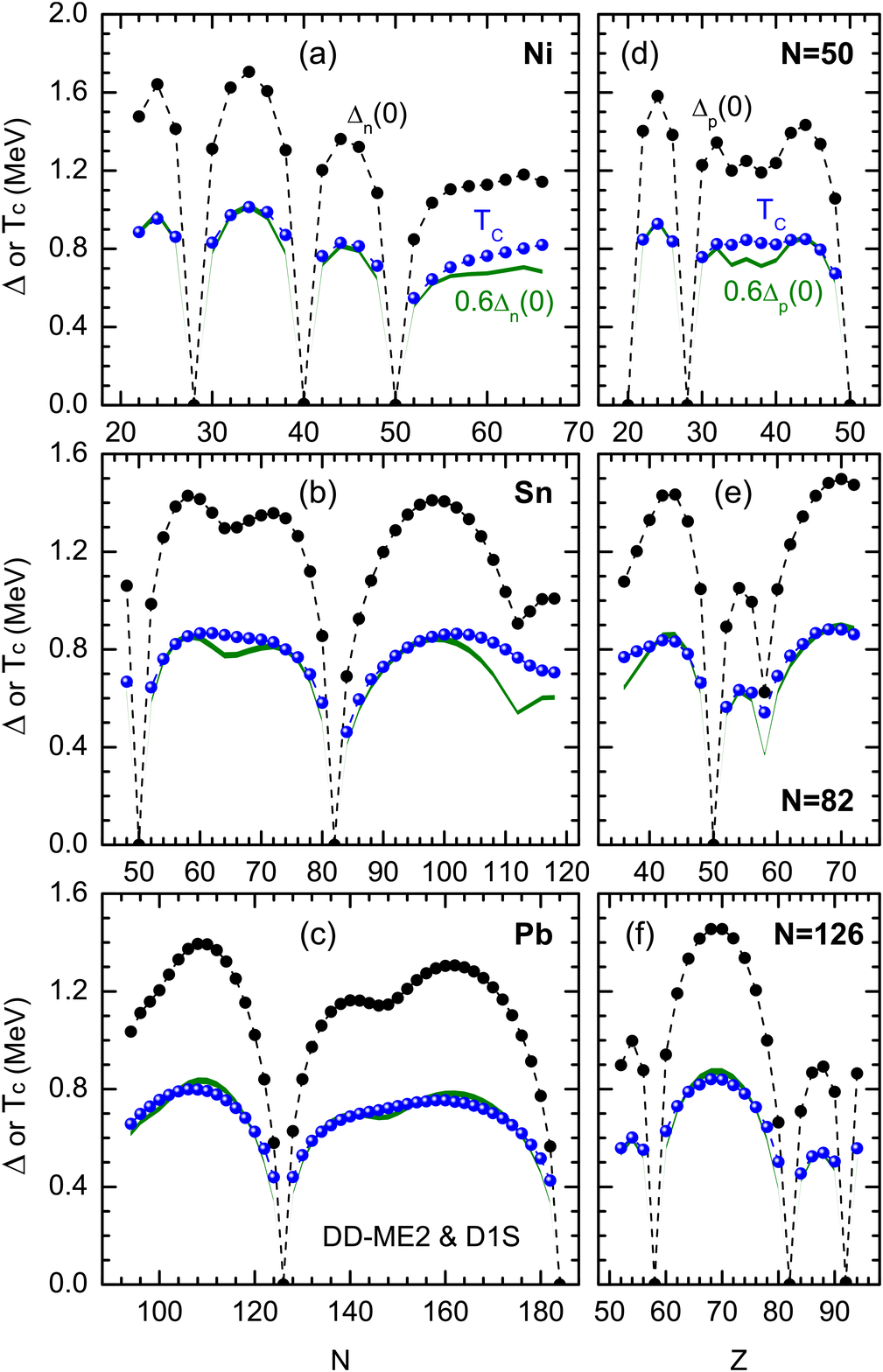}
%\fi
\caption{ (Color online)
The same as in Fig.~\ref{fig:GAPa1},
but calculated in the FT-RHB theory with the
effective interaction DD-ME2.
}
\label{fig:GAPme}
\end{figure}

We first discuss the pairing properties of finite nuclei at zero temperature, which are influenced by the underlying s.p. structure around the Fermi energy. For the Ni isotopes, a subshell closure at $N=40$ is predicted with PKA1 and PKO1 Lagrangians, as expected from  experiments~\cite{Sorlin2002}, while DD-ME2 shows a more pronounced shell closure. For neutron rich Ni isotopes, PKA1 indicates another subshell closure at $N=62$ which is not seen with PKO1 or DD-ME2. It is, however, beyond the present experimental limits. For the Sn isotopes, a decrease of the pairing gap induced by subshell closure is observed at $N=64$ with PKO1 and DD-ME2 Lagrangians, but not with PKA1. For Pb, a small decrease of the pairing gap is observed at $N=138$ with PKA1 and at $N=146$ with PKO1 and DD-ME2. On the isotonic side, we observe a subshell closure at $Z=40$ for $N=50$ with PKA1 Lagrangian, but not with PKO1 or DD-ME2. For $N=82$ isotones, PKA1 predicts a well marked shell closure at $Z=40$ and a subshell closure at $Z=34$ and $Z=64$, while at $Z=58$ PKO1 predicts a shell closure and DD-ME2 only a subshell closure. Finally, for $N=126$, PKO1 and DD-ME2 indicate a reduction of the pairing gap at $Z=92$, which is not confirmed by experimental data~\cite{Geng2006} and is not present with PKA1 Lagrangian. We will see below that these structures can have an impact on the thermal properties.

Turning now to the thermal properties of these isotopes and isotones, the comparison of the calculated critical
temperature $T_c$ and the approximate relation $0.60 \Delta(0)$ shown in Figs.~\ref{fig:GAPa1}-{\ref{fig:GAPme}
exhibits some interesting features.
The critical temperature $T_c$ and the approximate relation $0.60 \Delta(0)$ are identical in most cases with some exceptions.
In heavy nuclei (Pb and $N=126$), there are no strong deviations between these two quantities,
but they are however more marked in lighter nuclei.
Moreover, the cases where the exact and the approximate values of $T_c$ differ are correlated with either the presence of a subshell closure, or with the proximity of the drip lines.
In the case of subshell closure, the effect of the temperature is to "wash out" the decrease of the pairing
correlations.
This can be understood as the consequence of the thermal occupation probabilities which overcome small shell
gaps.
Close to the neutron drip lines, the more pronounced effects are observed in Ni and Sn isotopes.
This is also due to the thermal occupation of close-by resonant states as discussed in Refs.~\cite{Margueron2012a, Pastore2012}.
In the non-relativistic Skyrme Hartree-Fock plus BCS (SHF-BCS) approach, an enhancement of the critical temperature was
found in $^{140}$Sn using SkT6~\cite{Reib1999}.
We do not confirm this enhancement in $^{140}$Sn with the models used in this work. However, it
is interesting to notice that the origin of such an enhancement found in Ref.~\cite{Reib1999} is also related to
the existence of a subshell closure.

Let us finish this subsection with some general remarks concerning the nuclei which do not manifest any enhancement of
the critical temperature.
For the Pb isotopes, as shown in Figs.~\ref{fig:GAPa1}-\ref{fig:GAPme}, we have not observed any marked enhancement
of the critical temperature near the drip line as in the case of Sn isotopes.
Comparing Pb and Sn, since the pairing gap for these isotopes is decreasing near the drip line, one could
have expected to observe an enhancement of the critical temperature in Pb as it is observed in Sn.
For instance, the last occupied state in the drip line nucleus $^{266}$Pb  is indeed found to be well bound
($\epsilon_{3d_{3/2}} < -2.0$~MeV), and the lowest s.p. resonance $\epsilon_{2h_{11/2}}$ is found to be above 1.5 MeV.
There is therefore a rather strong gap in the neutron rich Pb isotopes ($N=184$) which
prevents the large coupling to the continuum.
For the isotonic chains, we do not find any pairing persistence phenomenon around the drip line.
This can be related to the well developed shell closures at $Z = 50, 82$ and 92 for proton-rich nuclei, which quench the
coupling to continuum states, as it would have been expected in such exotic nuclei.
In addition, the coupling to the continuum is weaker for protons since the Coulomb barrier tends to localize the
proton density in the nuclear interior~\cite{Dobaczewski1994}.
For these reasons, the persistence phenomenon is strongly quenched in the proton channel.

\subsection{Pairing persistence in $^{68}$Ni and $^{174}$Sn}

\begin{figure}[t]
\centering
%\ifpdf
%\includegraphics[width = 0.48\textwidth]{Ni68.pdf}
%\else
\includegraphics[width = 0.48\textwidth]{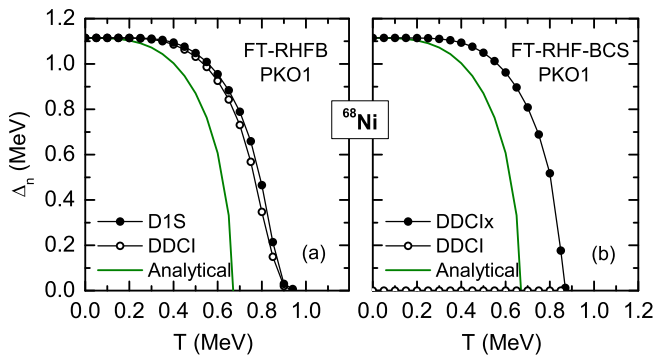}
%\fi
\caption{(Color online)
The neutron pairing gaps in $^{68}$Ni as a function of temperature, calculated by FT-RHFB with Gogny D1S and DDCI pairing forces,
and by FT-RHF-BCS with DDCI pairing force. To evaluate the persistence provoked by the subshell, the analytical results are also shown.
The pairing strength $V_0$ (in MeV.fm$^{-3}$) is fixed to be 326 (DDCI) and 537 (DDCIx).
}
\label{fig:Ni68}
\end{figure}

In this subsection, we analyze in more details the temperature dependence of the pairing gap for two representative nuclei, $^{68}$Ni and $^{174}$Sn. On the one hand, $^{68}$Ni is an isotope which is slightly more neutron rich than the five stable isotopes $^{58-64}$Ni. As shown in the previous subsection, $^{68}$Ni is considered as a subshell isotope~\cite{Kleban2002, Yamagami2012, Agbemava2014}, and as a consequence, the pairing gap at zero temperature is either reduced or strongly quenched depending on the model, see Figs.~\ref{fig:GAPa1}-{\ref{fig:GAPme}. On the other hand, $^{174}$Sn is a very neutron-rich isotope at or close to the neutron drip line, where the continuum effects are expected to be remarkable~\cite{Dobaczewski1996, Meng2002, Margueron2012a, Pastore2013}. However, since $^{174}$Sn is close to the potentially doubly magic $^{176}$Sn ($Z$=50, $N$=126), a gap is expected  to be present in the s.p. structure between bound and resonant states. These two nuclei are therefore representative of quantum systems for which pairing at zero temperature is weakened by the presence of a gap above the Fermi energy and where a finite amount of temperature allows to overcome the gap and provokes an enhancement of pairing correlations, giving rise to pairing persistence.

We first show the temperature dependence of the pairing gap for $^{68}$Ni in Fig.~\ref{fig:Ni68}(a), calculated with the FT-RHFB model where we consider the PKO1 Lagrangian in the mean field channel and either Gogny D1S or DDCI interaction in the pairing channel. The analytical model is also shown for reference. It is found that the predictions for $\Delta_n(T)$  do not practically depend on the pairing force. The critical temperature predicted by the FT-RHFB approach is increased with respect to the reference analytical model: the FT-RHFB pairing gaps vanish around $T = 0.90$~MeV, which is $0.25$~MeV higher than the expected value ($0.65$~MeV). The pairing gap predicted by FT-RHF-BCS is shown in Fig.~\ref{fig:Ni68}(b). Surprisingly, the pairing gap is zero if the same DDCI pairing interaction is used. An increase of the pairing strength $V_0$ is therefore necessary. It is also interesting to observe that the DDCIx pairing interaction, where $V_0$ is increased to match with the zero temperature pairing gap obtained with FT-RHFB case, reproduces almost exactly the temperature dependence of the FT-RHFB case and predicts as well an increase of the critical temperature with respect to the analytical model. The nucleus $^{68}$Ni is a typical example of pairing persistence at finite temperature in closed subshell ($N = 40$) nuclei. We hereafter name this phenomenon type I pairing persistence. Other examples of similar behavior are: $^{90}$Ni (PKA1), $^{114}$Sn (PKO1, DD-ME2), $^{220}$Pb (PKA1), $^{230}$Pb (DD-ME2) for the neutron pairing gap, and $^{90}$Zr (PKA1, PKO1 and DD-ME2), $^{140}$Ce (DD-ME2), $^{146, 190}$Gd (PKA1) for the proton pairing gap.

\begin{figure}[b]
\centering
%\ifpdf
%\includegraphics[width = 0.48\textwidth]{Sn174.pdf}
%\else
\includegraphics[width = 0.48\textwidth]{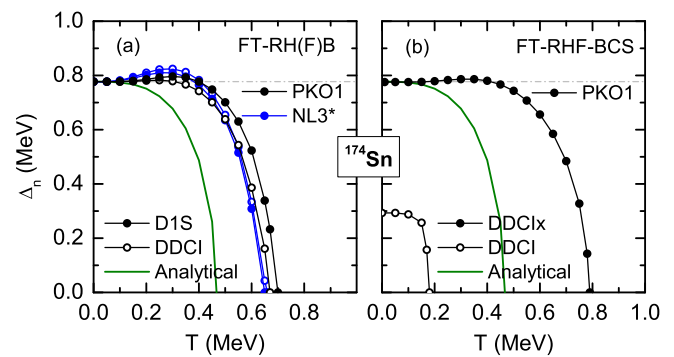}
%\fi
\caption{(Color online)
The neutron pairing gaps in $^{174}$Sn as a function of temperature,
calculated with PKO1 and NL3$^*$, using Gogny D1S and  DDCI pairing forces.
The results of the analytical model are also shown.
The pairing strength $V_0$ (in MeV.fm$^{-3}$) is: 333 (DDCI with PKO1), 317 (DDCI with NL3$^*$) and 596 (DDCIx with PKO1 and BCS framework).
}
\label{fig:Sn174}
\end{figure}

We turn now to the analysis of the results in $^{174}$Sn.
As stated above, this is a nucleus where pairing correlations are
slightly weakened due to the proximity of shell closure. A small amount of temperature is expected to reorganize the level occupancy around the
Fermi energy, opening more space below the Fermi energy, and producing a non-zero occupancy of the
first levels above the Fermi energy which are in the continuum.
Most of the occupied states in the continuum are resonance states, but it is interesting to notice that a small number
of them are also non-resonant states~\cite{Bennaceur1999}.
Without the participation of these non-resonant states, the asymptotic behavior of the density would be ill-defined and present
an unexpected gas component, as it is also observed in the BCS theory~\cite{Dobaczewski1996}.
The Bogoliubov transformation couples all states in a sub-$(l,j)$ space and a truncation among these states breaks the
unitarity of this transformation.
To avoid the presence of non-physical gas component in the density profile, it is therefore important to couple all states in the continuum within the Bogoliubov transformation.

Figure~\ref{fig:Sn174}(a) displays the evolution of the neutron pairing gap as a function of temperature in $^{174}$Sn with different Lagrangians and pairing forces. The results are very weakly affected by the choice of the pairing interaction. The effect induced by the choice of Lagrangian is also very small. We find a systematic increase of the critical temperature ($T_c \approx 0.65-0.70$~MeV) with respect to that expected from the analytical relation ($T_c \approx 0.47$~MeV), independently of the considered model. In addition, for temperature $T>0.2$~MeV, an increase of the pairing gap is observed, which can also be related to the thermally induced contribution of the continuum states. We compare these results to the ones obtained with the FT-RHF-BCS framework shown in Fig.~\ref{fig:Sn174}(b). As already discussed, the DDCI interaction predicts a reduced pairing gap in BCS compared to RHFB. On the other side, the DDCIx interaction where the pairing strength is increased to match the $T=0$ predictions of RHFB leads to an overestimation of the critical temperature compared to the FT-RHFB case. It shows that, in this case of dripline nucleus, the RHFB calculation cannot be simply reproduced by a BCS calculation where the pairing strength is just increased. Since the coupling to the continuum plays a dominant role in the persistence phenomenon in $^{174}$Sn, we hereafter call it type II phenomenon. From our results, it is also expected to occur  in Ni and Sn neutron-rich nuclei, Ni ($N>54\sim60$) and Sn ($N>100$).

\subsection{Entropy and specific heat}

\begin{figure}[t]
\centering
%\ifpdf
%\includegraphics[width = 0.48\textwidth]{Sn120.pdf}
%\else
\includegraphics[width = 0.48\textwidth]{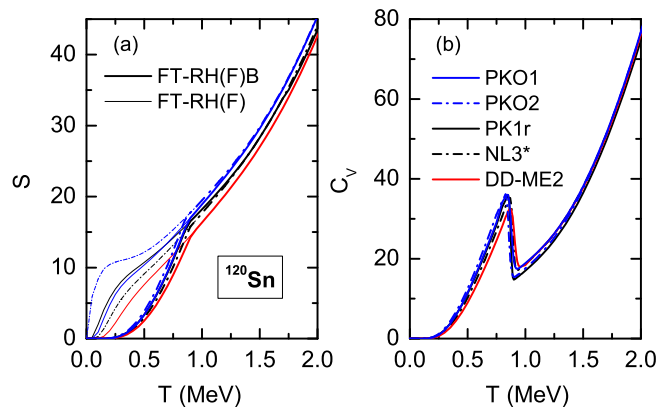}
%\fi
\caption{(Color online) Entropy and specific heat in $^{120}$Sn as a function of temperature, calculated using the FT-RH(F)B and FT-RH(F) theories.}
\label{fig:Sn120}
\end{figure}

We now focus on the entropy $S$ and specific heat $C_v$ which are the first and second derivatives of the free energy $F$ with respect to the temperature, and thus sensitive to thermal changes of the ground state, see for instance Ref.~\cite{Pastore2015} and Refs. therein. To test the sensitivity of these quantities to the choice of different models, we select two Sn isotopes, $^{120}$Sn and $^{160}$Sn. The former is a good example of a stable well-paired nucleus, while the latter is yet far from actual nuclear experiments but represents an extreme case with large isospin asymmetry.

In Fig.~\ref{fig:Sn120} are shown the entropy $S$ and the specific heat $C_v$ as functions of the temperature calculated by the RHF functionals PKO1 and PKO2, the RH ones with the non-linear self-couplings PK1r and NL3$^*$, and the RH one with density-dependent meson-nucleon couplings DD-ME2}. In the pairing channel, the value of the scaling factor $g$ (see first paragraph of section~\ref{results}), is slightly modified to give identical pairing gaps at zero temperature for the different models. In Fig.~\ref{fig:Sn120}(a), the entropy calculated with the pairing correlations [labelled by FT-RH(F)B] is compared to that neglecting the pairing correlations [labelled by FT-RH(F)]. At low temperature, if the pairing effects are ignored, the entropy is found to be largely model-dependent, i.e., the model with smaller non-relativistic effective mass (see Table~\ref{tab:NMP}), which leads to larger s.p. level spacing on the average, presents smaller entropy. As the temperature increases, and also as the pairing correlations are switched on, the entropy becomes less model-dependent.

In fact, at low temperature or without pairing correlations, the entropy is largely determined by the few states around the Fermi energy, and the number of the involved states is essentially determined by the detailed s.p. spectrum which depends on the models, therefore leading to model-dependent entropy. Both temperature and pairing correlations can disperse the particle over the states beyond the Fermi level. As the temperature increases, and/or as the pairing correlations are enhanced, more s.p. states will get involved to contribute to the entropy, and the average properties such as the density of states will become dominant, instead of a few states in the FT-RH(F) cases. Compared to distinctly different s.p. spectra around the Fermi surface, the dispersions of the non-relativistic effective masses (see Table~\ref{tab:NMP}) between the models are less remarkable. Even though, in Fig.~\ref{fig:Sn120}(a) it is clearly shown that the FT-RH(F)B results are grouped by the values of the effective masses when $T\gtrsim1$~MeV which correspond to different average densities of states. As expected, the effect of the pairing correlations is clearly visible below the critical temperature ($T_c\approx 0.8$~MeV), inducing a strong reduction of the entropy [see Fig.~\ref{fig:Sn120}(a)] and singular behaviors of the specific heat around the critical temperature as shown in Fig.~\ref{fig:Sn120}(b). Just above the critical temperature, we can notice that the specific heat is not linear in $T$, as expected from the Fermi gas model~\cite{Gilbert1965, Schiller2001}, and the linear dependence seems to be found at slightly larger temperature ($T>1.5$~MeV). The non-linearity of the specific heat around $T_c$ might be related to shell effects.

The results thus clearly show that the pairing correlations contribute to the $s$-shaped behavior of the specific
heat, as it has already been inferred from the analysis of thermal excited nuclei in laboratory
experiments~\cite{Schiller2001, Guttormsen2001, Agvaanluvsan2009}.
A realistic description of the smooth $s$-shaped behavior in finite nuclei requires a more elaborated modelling,
including for instance particle number projection~\cite{Tanabe1981, Esebbag1993, Liu2001, Dinh2003}.
It is however shown in Ref.~\cite{Gambacurta2013} that the smooth $s$-shaped behavior may be even washed out in some rare earth nuclei. The results presented in Fig.~\ref{fig:Sn120} should not be compared directly to the semi-experimental data.

\begin{figure}[t]
\centering
%\ifpdf
%\includegraphics[width = 0.48\textwidth]{Sn160.pdf}
%\else
\includegraphics[width = 0.48\textwidth]{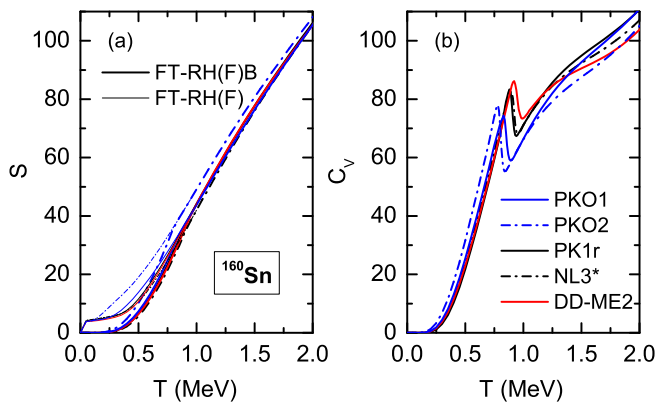}
%\fi
\caption{(Color online) The same as in Fig.~\ref{fig:Sn120}, but for $^{160}$Sn.}
\label{fig:Sn160}
\end{figure}

The situation for the neutron-rich nucleus $^{160}$Sn is more complex as shown in Fig.~\ref{fig:Sn160}.
The model dependence of entropy at low temperature is reduced compared to the case of $^{120}$Sn.
Only PKO2 predictions differ from the other models.
The predictions for the critical temperature, associated to the discontinuity in the specific heat of Fig.~\ref{fig:Sn160},
vary among the modellings to a much larger extent than what was found in $^{120}$Sn.
$^{160}$Sn is located in the region where pairing persistence is expected to appear, see
Figs.~\ref{fig:GAPa1}-{\ref{fig:GAPme}. Since this phenomenon is strongly related to the position of
resonance states in the continuum, we expect to observe deviations among models predicting different
positions of these states.
This model dependence therefore reveals our lack of knowledge in extrapolating Lagrangians which have been adjusted
for less exotic nuclei.

\section{SUMMARY and CONCLUSION}\label{summary}

In this work, we have developed the first FT-RHFB theory for spherical nuclei.
The self-consistent FT-RHFB equations are solved by using a DWS basis which provides
an appropriate asymptotic behavior for the continuum states.
We have performed systematic FT-RHFB calculations for both stable and weakly bound nuclei and discussed
their thermal properties.
The influence of the pairing interaction on the pairing phase transition is evaluated.
It is found that the critical temperature for a pairing transition generally follows the rule
$T_c = 0.60\Delta(0)$ with a finite-range pairing force and $T_c = 0.57\Delta(0)$ with a contact pairing force.
The finite- or zero-range nature of the pairing force, while generating different state-dependence pairing gaps,
causes only small differences in our results.
We have described the pairing persistence in two kinds of situations: nuclei at subshell closure (type I),
and nuclei strongly coupled to continuum states which are close to the drip line (type II).
We have observed that, while a refitting of the pairing strength could match the FT-RHF-BCS with the FT-RHFB
predictions for the pairing gap in the case I, it is no longer the true in the case II.
This is due to the participation of the continuum states in the second case which involve coupling of different nature.

We have also analyzed the influence of the interaction on the thermal response.
The results show clearly that the pairing correlations contribute to the $s$-shaped behavior of the specific heat curve,
and help to wash out the model dependence.
For stable nuclei the model deviations, to some extent, can be traced back to the effective mass, since
the level structure only weakly depends on the choice of the CDF.
The situation for exotic nuclei is more complex since it is related to our lack of knowledge in very exotic nuclei,
and the pairing persistence would have large effects on their thermal property.

In conclusion, we have illustrated the richness and complexity of pairing correlations at finite temperature and in finite systems within the first FT-RHFB calculation.
The discussion of correlations beyond mean-field, induced for instance by particle number projection, is not addressed in this work.
It is however expected that the particle number projection will contribute to increase the pairing correlations in the case where they are weak~\cite{Gambacurta2013},
like in the pairing persistence phenomenon discussed in this paper.
In future work, a more quantitative calculation will be necessary to estimate how strong are these additional correlations and how they modify the results presented in this work.
Another interesting perspective which is suggested by this work is the possibility that similar phenomena can be observed in other domains of physics. For instance, it was studied whether cold atoms in a double potential could demonstrate pairing persistence as well~\cite{Pastore2014}.
Finally, the application of this formalism for the prediction of temperature evolution of pairing properties in the crust of neutron stars~\cite{Sandulescu2004, Grill2011, Margueron2012b} will be performed in the near future.
There, the thermal modification of pairing correlations could have a large impact on the thermal relaxation of the crust~\cite{Fortin2010}, and could be observed during the quiescent period of low mass X-ray transients.

\begin{acknowledgements}
J. Li acknowledges the support by China Scholarship Council (CSC) under grant No. 201206180063.
This work was partially supported by the NewCompStar COST action and the ANR project SN2NS No. ANR-10-BLAN-0503.
Research at LZU was supported by the National Natural Science Foundation of China under grants No. 11375076 and No. 11405223, and by the Specialized Research Fund for the Doctoral Program of Higher Education under grant No. 20130211110005.
\end{acknowledgements}

\bibliographystyle{apsrev}
\bibliography{references}	

\begin{thebibliography}{92}
\expandafter\ifx\csname natexlab\endcsname\relax\def\natexlab#1{#1}\fi
\expandafter\ifx\csname bibnamefont\endcsname\relax
  \def\bibnamefont#1{#1}\fi
\expandafter\ifx\csname bibfnamefont\endcsname\relax
  \def\bibfnamefont#1{#1}\fi
\expandafter\ifx\csname citenamefont\endcsname\relax
  \def\citenamefont#1{#1}\fi
\expandafter\ifx\csname url\endcsname\relax
  \def\url#1{\texttt{#1}}\fi
\expandafter\ifx\csname urlprefix\endcsname\relax\def\urlprefix{URL }\fi
\providecommand{\bibinfo}[2]{#2}
\providecommand{\eprint}[2][]{\url{#2}}

\bibitem[{\citenamefont{Melby et~al.}(1999)\citenamefont{Melby, Bergholt,
  Guttormsen, Hjorth-Jensen, Ingebretsen, Messelt, Rekstad, Schiller, Siem, and
  \O{}deg\aa{}rd}}]{Melby1999}
\bibinfo{author}{\bibfnamefont{E.}~\bibnamefont{Melby}},
  \bibinfo{author}{\bibfnamefont{L.}~\bibnamefont{Bergholt}},
  \bibinfo{author}{\bibfnamefont{M.}~\bibnamefont{Guttormsen}},
  \bibinfo{author}{\bibfnamefont{M.}~\bibnamefont{Hjorth-Jensen}},
  \bibinfo{author}{\bibfnamefont{F.}~\bibnamefont{Ingebretsen}},
  \bibinfo{author}{\bibfnamefont{S.}~\bibnamefont{Messelt}},
  \bibinfo{author}{\bibfnamefont{J.}~\bibnamefont{Rekstad}},
  \bibinfo{author}{\bibfnamefont{A.}~\bibnamefont{Schiller}},
  \bibinfo{author}{\bibfnamefont{S.}~\bibnamefont{Siem}}, \bibnamefont{and}
  \bibinfo{author}{\bibfnamefont{S.~W.} \bibnamefont{\O{}deg\aa{}rd}},
  \bibinfo{journal}{Phys. Rev. Lett.} \textbf{\bibinfo{volume}{83}},
  \bibinfo{pages}{3150} (\bibinfo{year}{1999}).

\bibitem[{\citenamefont{Agvaanluvsan et~al.}(2004)\citenamefont{Agvaanluvsan,
  Schiller, Becker, Bernstein, Garrett, Guttormsen, Mitchell, Rekstad, Siem,
  Voinov et~al.}}]{Agvaanluvsan2004}
\bibinfo{author}{\bibfnamefont{U.}~\bibnamefont{Agvaanluvsan}},
  \bibinfo{author}{\bibfnamefont{A.}~\bibnamefont{Schiller}},
  \bibinfo{author}{\bibfnamefont{J.~A.} \bibnamefont{Becker}},
  \bibinfo{author}{\bibfnamefont{L.~A.} \bibnamefont{Bernstein}},
  \bibinfo{author}{\bibfnamefont{P.~E.} \bibnamefont{Garrett}},
  \bibinfo{author}{\bibfnamefont{M.}~\bibnamefont{Guttormsen}},
  \bibinfo{author}{\bibfnamefont{G.~E.} \bibnamefont{Mitchell}},
  \bibinfo{author}{\bibfnamefont{J.}~\bibnamefont{Rekstad}},
  \bibinfo{author}{\bibfnamefont{S.}~\bibnamefont{Siem}},
  \bibinfo{author}{\bibfnamefont{A.}~\bibnamefont{Voinov}},
  \bibnamefont{et~al.}, \bibinfo{journal}{Phys. Rev. C}
  \textbf{\bibinfo{volume}{70}}, \bibinfo{pages}{054611}
  (\bibinfo{year}{2004}).

\bibitem[{\citenamefont{Kalmykov et~al.}(2007)\citenamefont{Kalmykov, \"Ozen,
  Langanke, Mart\'inez-Pinedo, von Neumann-Cosel, and Richter}}]{Kalmykov2007}
\bibinfo{author}{\bibfnamefont{Y.}~\bibnamefont{Kalmykov}},
  \bibinfo{author}{\bibfnamefont{C.}~\bibnamefont{\"Ozen}},
  \bibinfo{author}{\bibfnamefont{K.}~\bibnamefont{Langanke}},
  \bibinfo{author}{\bibfnamefont{G.}~\bibnamefont{Mart\'inez-Pinedo}},
  \bibinfo{author}{\bibfnamefont{P.}~\bibnamefont{von Neumann-Cosel}},
  \bibnamefont{and} \bibinfo{author}{\bibfnamefont{A.}~\bibnamefont{Richter}},
  \bibinfo{journal}{Phys. Rev. Lett.} \textbf{\bibinfo{volume}{99}},
  \bibinfo{pages}{202502} (\bibinfo{year}{2007}).

\bibitem[{\citenamefont{Toft et~al.}(2010)\citenamefont{Toft, Larsen,
  Agvaanluvsan, B\"urger, Guttormsen, Mitchell, Nyhus, Schiller, Siem, Syed
  et~al.}}]{Toft2010}
\bibinfo{author}{\bibfnamefont{H.~K.} \bibnamefont{Toft}},
  \bibinfo{author}{\bibfnamefont{A.~C.} \bibnamefont{Larsen}},
  \bibinfo{author}{\bibfnamefont{U.}~\bibnamefont{Agvaanluvsan}},
  \bibinfo{author}{\bibfnamefont{A.}~\bibnamefont{B\"urger}},
  \bibinfo{author}{\bibfnamefont{M.}~\bibnamefont{Guttormsen}},
  \bibinfo{author}{\bibfnamefont{G.~E.} \bibnamefont{Mitchell}},
  \bibinfo{author}{\bibfnamefont{H.~T.} \bibnamefont{Nyhus}},
  \bibinfo{author}{\bibfnamefont{A.}~\bibnamefont{Schiller}},
  \bibinfo{author}{\bibfnamefont{S.}~\bibnamefont{Siem}},
  \bibinfo{author}{\bibfnamefont{N.~U.~H.} \bibnamefont{Syed}},
  \bibnamefont{et~al.}, \bibinfo{journal}{Phys. Rev. C}
  \textbf{\bibinfo{volume}{81}}, \bibinfo{pages}{064311}
  (\bibinfo{year}{2010}).

\bibitem[{\citenamefont{Guttormsen et~al.}(2014)\citenamefont{Guttormsen,
  Larsen, Garrote, Byun, Eriksen, Giacoppo, G\"orgen, Hagen, Klintefjord, Nyhus
  et~al.}}]{Guttormsen2014}
\bibinfo{author}{\bibfnamefont{M.}~\bibnamefont{Guttormsen}},
  \bibinfo{author}{\bibfnamefont{A.~C.} \bibnamefont{Larsen}},
  \bibinfo{author}{\bibfnamefont{F.~L.~B.} \bibnamefont{Garrote}},
  \bibinfo{author}{\bibfnamefont{Y.}~\bibnamefont{Byun}},
  \bibinfo{author}{\bibfnamefont{T.~K.} \bibnamefont{Eriksen}},
  \bibinfo{author}{\bibfnamefont{F.}~\bibnamefont{Giacoppo}},
  \bibinfo{author}{\bibfnamefont{A.}~\bibnamefont{G\"orgen}},
  \bibinfo{author}{\bibfnamefont{T.~W.} \bibnamefont{Hagen}},
  \bibinfo{author}{\bibfnamefont{M.}~\bibnamefont{Klintefjord}},
  \bibinfo{author}{\bibfnamefont{H.~T.} \bibnamefont{Nyhus}},
  \bibnamefont{et~al.}, \bibinfo{journal}{Phys. Rev. C}
  \textbf{\bibinfo{volume}{90}}, \bibinfo{pages}{044309}
  (\bibinfo{year}{2014}).

\bibitem[{\citenamefont{Balian et~al.}(1999)\citenamefont{Balian, Flocard, and
  V\'en\'eroni}}]{Balian1999}
\bibinfo{author}{\bibfnamefont{R.}~\bibnamefont{Balian}},
  \bibinfo{author}{\bibfnamefont{H.}~\bibnamefont{Flocard}}, \bibnamefont{and}
  \bibinfo{author}{\bibfnamefont{M.}~\bibnamefont{V\'en\'eroni}},
  \bibinfo{journal}{Phys. Rep.} \textbf{\bibinfo{volume}{317}},
  \bibinfo{pages}{251} (\bibinfo{year}{1999}).

\bibitem[{\citenamefont{Liu and Alhassid}(2001)}]{Liu2001}
\bibinfo{author}{\bibfnamefont{S.}~\bibnamefont{Liu}} \bibnamefont{and}
  \bibinfo{author}{\bibfnamefont{Y.}~\bibnamefont{Alhassid}},
  \bibinfo{journal}{Phys. Rev. Lett.} \textbf{\bibinfo{volume}{87}},
  \bibinfo{pages}{022501} (\bibinfo{year}{2001}).

\bibitem[{\citenamefont{Guttormsen et~al.}(2001)\citenamefont{Guttormsen,
  Hjorth-Jensen, Melby, Rekstad, Schiller, and Siem}}]{Guttormsen2001}
\bibinfo{author}{\bibfnamefont{M.}~\bibnamefont{Guttormsen}},
  \bibinfo{author}{\bibfnamefont{M.}~\bibnamefont{Hjorth-Jensen}},
  \bibinfo{author}{\bibfnamefont{E.}~\bibnamefont{Melby}},
  \bibinfo{author}{\bibfnamefont{J.}~\bibnamefont{Rekstad}},
  \bibinfo{author}{\bibfnamefont{A.}~\bibnamefont{Schiller}}, \bibnamefont{and}
  \bibinfo{author}{\bibfnamefont{S.}~\bibnamefont{Siem}},
  \bibinfo{journal}{Phys. Rev. C} \textbf{\bibinfo{volume}{64}},
  \bibinfo{pages}{034319} (\bibinfo{year}{2001}).

\bibitem[{\citenamefont{Dean and Hjorth-Jensen}(2003)}]{Dean2003}
\bibinfo{author}{\bibfnamefont{D.~J.} \bibnamefont{Dean}} \bibnamefont{and}
  \bibinfo{author}{\bibfnamefont{M.}~\bibnamefont{Hjorth-Jensen}},
  \bibinfo{journal}{Rev. Mod. Phys.} \textbf{\bibinfo{volume}{75}},
  \bibinfo{pages}{607} (\bibinfo{year}{2003}).

\bibitem[{\citenamefont{Agvaanluvsan et~al.}(2009)\citenamefont{Agvaanluvsan,
  Larsen, Guttormsen, Chankova, Mitchell, Schiller, Siem, and
  Voinov}}]{Agvaanluvsan2009}
\bibinfo{author}{\bibfnamefont{U.}~\bibnamefont{Agvaanluvsan}},
  \bibinfo{author}{\bibfnamefont{A.~C.} \bibnamefont{Larsen}},
  \bibinfo{author}{\bibfnamefont{M.}~\bibnamefont{Guttormsen}},
  \bibinfo{author}{\bibfnamefont{R.}~\bibnamefont{Chankova}},
  \bibinfo{author}{\bibfnamefont{G.~E.} \bibnamefont{Mitchell}},
  \bibinfo{author}{\bibfnamefont{A.}~\bibnamefont{Schiller}},
  \bibinfo{author}{\bibfnamefont{S.}~\bibnamefont{Siem}}, \bibnamefont{and}
  \bibinfo{author}{\bibfnamefont{A.}~\bibnamefont{Voinov}},
  \bibinfo{journal}{Phys. Rev. C} \textbf{\bibinfo{volume}{79}},
  \bibinfo{pages}{014320} (\bibinfo{year}{2009}).

\bibitem[{\citenamefont{Sandulescu}(2004)}]{Sandulescu2004}
\bibinfo{author}{\bibfnamefont{N.}~\bibnamefont{Sandulescu}},
  \bibinfo{journal}{Phys. Rev. C} \textbf{\bibinfo{volume}{70}},
  \bibinfo{pages}{025801} (\bibinfo{year}{2004}).

\bibitem[{\citenamefont{Fortin et~al.}(2010)\citenamefont{Fortin, Grill,
  Margueron, Page, and Sandulescu}}]{Fortin2010}
\bibinfo{author}{\bibfnamefont{M.}~\bibnamefont{Fortin}},
  \bibinfo{author}{\bibfnamefont{F.}~\bibnamefont{Grill}},
  \bibinfo{author}{\bibfnamefont{J.}~\bibnamefont{Margueron}},
  \bibinfo{author}{\bibfnamefont{D.}~\bibnamefont{Page}}, \bibnamefont{and}
  \bibinfo{author}{\bibfnamefont{N.}~\bibnamefont{Sandulescu}},
  \bibinfo{journal}{Phys. Rev. C} \textbf{\bibinfo{volume}{82}},
  \bibinfo{pages}{065804} (\bibinfo{year}{2010}).

\bibitem[{\citenamefont{Pastore et~al.}(2011)\citenamefont{Pastore, Baroni, and
  Losa}}]{Pastore2011}
\bibinfo{author}{\bibfnamefont{A.}~\bibnamefont{Pastore}},
  \bibinfo{author}{\bibfnamefont{S.}~\bibnamefont{Baroni}}, \bibnamefont{and}
  \bibinfo{author}{\bibfnamefont{C.}~\bibnamefont{Losa}},
  \bibinfo{journal}{Phys. Rev. C} \textbf{\bibinfo{volume}{84}},
  \bibinfo{pages}{065807} (\bibinfo{year}{2011}).

\bibitem[{\citenamefont{Margueron and Khan}(2012)}]{Margueron2012a}
\bibinfo{author}{\bibfnamefont{J.}~\bibnamefont{Margueron}} \bibnamefont{and}
  \bibinfo{author}{\bibfnamefont{E.}~\bibnamefont{Khan}},
  \bibinfo{journal}{Phys. Rev. C} \textbf{\bibinfo{volume}{86}},
  \bibinfo{pages}{065801} (\bibinfo{year}{2012}).

\bibitem[{\citenamefont{Pastore et~al.}(2013)\citenamefont{Pastore, Margueron,
  Schuck, and Vi\~nas}}]{Pastore2013}
\bibinfo{author}{\bibfnamefont{A.}~\bibnamefont{Pastore}},
  \bibinfo{author}{\bibfnamefont{J.}~\bibnamefont{Margueron}},
  \bibinfo{author}{\bibfnamefont{P.}~\bibnamefont{Schuck}}, \bibnamefont{and}
  \bibinfo{author}{\bibfnamefont{X.}~\bibnamefont{Vi\~nas}},
  \bibinfo{journal}{Phys. Rev. C} \textbf{\bibinfo{volume}{88}},
  \bibinfo{pages}{034314} (\bibinfo{year}{2013}).

\bibitem[{\citenamefont{Sedrakian et~al.}(1997)\citenamefont{Sedrakian, Alm,
  and Lombardo}}]{Sedrakian1997}
\bibinfo{author}{\bibfnamefont{A.}~\bibnamefont{Sedrakian}},
  \bibinfo{author}{\bibfnamefont{T.}~\bibnamefont{Alm}}, \bibnamefont{and}
  \bibinfo{author}{\bibfnamefont{U.}~\bibnamefont{Lombardo}},
  \bibinfo{journal}{Phys. Rev. C} \textbf{\bibinfo{volume}{55}},
  \bibinfo{pages}{R582} (\bibinfo{year}{1997}).

\bibitem[{\citenamefont{Hung and Dang}(2011)}]{Hung2011}
\bibinfo{author}{\bibfnamefont{N.~Q.} \bibnamefont{Hung}} \bibnamefont{and}
  \bibinfo{author}{\bibfnamefont{N.~D.} \bibnamefont{Dang}},
  \bibinfo{journal}{Phys. Rev. C} \textbf{\bibinfo{volume}{84}},
  \bibinfo{pages}{054324} (\bibinfo{year}{2011}).

\bibitem[{\citenamefont{Sano and Yamasaki}(1963)}]{Sano1963}
\bibinfo{author}{\bibfnamefont{M.}~\bibnamefont{Sano}} \bibnamefont{and}
  \bibinfo{author}{\bibfnamefont{S.}~\bibnamefont{Yamasaki}},
  \bibinfo{journal}{Prog. Theor. Phys.} \textbf{\bibinfo{volume}{29}},
  \bibinfo{pages}{397} (\bibinfo{year}{1963}).

\bibitem[{\citenamefont{Goodman}(1981)}]{Goodman1981}
\bibinfo{author}{\bibfnamefont{A.~L.} \bibnamefont{Goodman}},
  \bibinfo{journal}{Nucl. Phys. A} \textbf{\bibinfo{volume}{352}},
  \bibinfo{pages}{30} (\bibinfo{year}{1981}).

\bibitem[{\citenamefont{Goodman}(1986)}]{Goodman1986}
\bibinfo{author}{\bibfnamefont{A.~L.} \bibnamefont{Goodman}},
  \bibinfo{journal}{Phys. Rev. C} \textbf{\bibinfo{volume}{34}},
  \bibinfo{pages}{1942} (\bibinfo{year}{1986}).

\bibitem[{\citenamefont{Egido and Ring}(1993)}]{Egido1993}
\bibinfo{author}{\bibfnamefont{J.~L.} \bibnamefont{Egido}} \bibnamefont{and}
  \bibinfo{author}{\bibfnamefont{P.}~\bibnamefont{Ring}}, \bibinfo{journal}{J.
  Phys G: Nucl. Part. Phys.} \textbf{\bibinfo{volume}{19}}, \bibinfo{pages}{1}
  (\bibinfo{year}{1993}).

\bibitem[{\citenamefont{Rei{\ss} et~al.}(1999)\citenamefont{Rei{\ss}, Bender,
  and Reinhard}}]{Reib1999}
\bibinfo{author}{\bibfnamefont{C.}~\bibnamefont{Rei{\ss}}},
  \bibinfo{author}{\bibfnamefont{M.}~\bibnamefont{Bender}}, \bibnamefont{and}
  \bibinfo{author}{\bibfnamefont{P.-G.} \bibnamefont{Reinhard}},
  \bibinfo{journal}{Eur. Phys. J. A} \textbf{\bibinfo{volume}{6}},
  \bibinfo{pages}{157} (\bibinfo{year}{1999}).

\bibitem[{\citenamefont{Khan et~al.}(2007)\citenamefont{Khan, Van~Giai, and
  Sandulescu}}]{Khan2007}
\bibinfo{author}{\bibfnamefont{E.}~\bibnamefont{Khan}},
  \bibinfo{author}{\bibfnamefont{N.}~\bibnamefont{Van~Giai}}, \bibnamefont{and}
  \bibinfo{author}{\bibfnamefont{N.}~\bibnamefont{Sandulescu}},
  \bibinfo{journal}{Nucl. Phys. A} \textbf{\bibinfo{volume}{789}},
  \bibinfo{pages}{94} (\bibinfo{year}{2007}).

\bibitem[{\citenamefont{Niu et~al.}(2013)\citenamefont{Niu, Niu, Paar,
  Vretenar, Wang, Bai, and Meng}}]{Niu2013}
\bibinfo{author}{\bibfnamefont{Y.~F.} \bibnamefont{Niu}},
  \bibinfo{author}{\bibfnamefont{Z.~M.} \bibnamefont{Niu}},
  \bibinfo{author}{\bibfnamefont{N.}~\bibnamefont{Paar}},
  \bibinfo{author}{\bibfnamefont{D.}~\bibnamefont{Vretenar}},
  \bibinfo{author}{\bibfnamefont{G.~H.} \bibnamefont{Wang}},
  \bibinfo{author}{\bibfnamefont{J.~S.} \bibnamefont{Bai}}, \bibnamefont{and}
  \bibinfo{author}{\bibfnamefont{J.}~\bibnamefont{Meng}},
  \bibinfo{journal}{Phys. Rev. C} \textbf{\bibinfo{volume}{88}},
  \bibinfo{pages}{034308} (\bibinfo{year}{2013}).

\bibitem[{\citenamefont{Gambhir et~al.}(2000)\citenamefont{Gambhir, Maharana,
  Lalazissis, Panos, and Ring}}]{Gambhir2000}
\bibinfo{author}{\bibfnamefont{Y.~K.} \bibnamefont{Gambhir}},
  \bibinfo{author}{\bibfnamefont{J.~P.} \bibnamefont{Maharana}},
  \bibinfo{author}{\bibfnamefont{G.~A.} \bibnamefont{Lalazissis}},
  \bibinfo{author}{\bibfnamefont{C.~P.} \bibnamefont{Panos}}, \bibnamefont{and}
  \bibinfo{author}{\bibfnamefont{P.}~\bibnamefont{Ring}},
  \bibinfo{journal}{Phys. Rev. C} \textbf{\bibinfo{volume}{62}},
  \bibinfo{pages}{054610} (\bibinfo{year}{2000}).

\bibitem[{\citenamefont{Egido et~al.}(2000)\citenamefont{Egido, Robledo, and
  Martin}}]{Egido2000}
\bibinfo{author}{\bibfnamefont{J.~L.} \bibnamefont{Egido}},
  \bibinfo{author}{\bibfnamefont{L.~M.} \bibnamefont{Robledo}},
  \bibnamefont{and} \bibinfo{author}{\bibfnamefont{V.}~\bibnamefont{Martin}},
  \bibinfo{journal}{Phys. Rev. Lett.} \textbf{\bibinfo{volume}{85}},
  \bibinfo{pages}{26} (\bibinfo{year}{2000}).

\bibitem[{\citenamefont{Agrawal et~al.}(2001)\citenamefont{Agrawal, Sil,
  Samaddar, and De}}]{Agrawal2001}
\bibinfo{author}{\bibfnamefont{B.~K.} \bibnamefont{Agrawal}},
  \bibinfo{author}{\bibfnamefont{T.}~\bibnamefont{Sil}},
  \bibinfo{author}{\bibfnamefont{S.~K.} \bibnamefont{Samaddar}},
  \bibnamefont{and} \bibinfo{author}{\bibfnamefont{J.~N.} \bibnamefont{De}},
  \bibinfo{journal}{Phys. Rev. C} \textbf{\bibinfo{volume}{63}},
  \bibinfo{pages}{024002} (\bibinfo{year}{2001}).

\bibitem[{\citenamefont{Martin et~al.}(2003)\citenamefont{Martin, Egido, and
  Robledo}}]{Martin2003}
\bibinfo{author}{\bibfnamefont{V.}~\bibnamefont{Martin}},
  \bibinfo{author}{\bibfnamefont{J.~L.} \bibnamefont{Egido}}, \bibnamefont{and}
  \bibinfo{author}{\bibfnamefont{L.~M.} \bibnamefont{Robledo}},
  \bibinfo{journal}{Phys. Rev. C} \textbf{\bibinfo{volume}{68}},
  \bibinfo{pages}{034327} (\bibinfo{year}{2003}).

\bibitem[{\citenamefont{Gambacurta et~al.}(2013)\citenamefont{Gambacurta,
  Lacroix, and Sandulescu}}]{Gambacurta2013}
\bibinfo{author}{\bibfnamefont{D.}~\bibnamefont{Gambacurta}},
  \bibinfo{author}{\bibfnamefont{D.}~\bibnamefont{Lacroix}}, \bibnamefont{and}
  \bibinfo{author}{\bibfnamefont{N.}~\bibnamefont{Sandulescu}},
  \bibinfo{journal}{Phys. Rev. C} \textbf{\bibinfo{volume}{88}},
  \bibinfo{pages}{034324} (\bibinfo{year}{2013}).

\bibitem[{\citenamefont{Dobaczewski et~al.}(1984)\citenamefont{Dobaczewski,
  Flocard, and Treiner}}]{Dobaczewski1984}
\bibinfo{author}{\bibfnamefont{J.}~\bibnamefont{Dobaczewski}},
  \bibinfo{author}{\bibfnamefont{H.}~\bibnamefont{Flocard}}, \bibnamefont{and}
  \bibinfo{author}{\bibfnamefont{J.}~\bibnamefont{Treiner}},
  \bibinfo{journal}{Nucl. Phys. A} \textbf{\bibinfo{volume}{422}},
  \bibinfo{pages}{103} (\bibinfo{year}{1984}).

\bibitem[{\citenamefont{Stoitsov et~al.}(1998)\citenamefont{Stoitsov, Ring,
  Vretenar, and Lalazissis}}]{Stoitsov1998}
\bibinfo{author}{\bibfnamefont{M.}~\bibnamefont{Stoitsov}},
  \bibinfo{author}{\bibfnamefont{P.}~\bibnamefont{Ring}},
  \bibinfo{author}{\bibfnamefont{D.}~\bibnamefont{Vretenar}}, \bibnamefont{and}
  \bibinfo{author}{\bibfnamefont{G.~A.} \bibnamefont{Lalazissis}},
  \bibinfo{journal}{Phys. Rev. C} \textbf{\bibinfo{volume}{58}},
  \bibinfo{pages}{2086} (\bibinfo{year}{1998}).

\bibitem[{\citenamefont{Zhou et~al.}(2003)\citenamefont{Zhou, Meng, and
  Ring}}]{Zhou2003}
\bibinfo{author}{\bibfnamefont{S.-G.} \bibnamefont{Zhou}},
  \bibinfo{author}{\bibfnamefont{J.}~\bibnamefont{Meng}}, \bibnamefont{and}
  \bibinfo{author}{\bibfnamefont{P.}~\bibnamefont{Ring}},
  \bibinfo{journal}{Phys. Rev. C} \textbf{\bibinfo{volume}{68}},
  \bibinfo{pages}{034323} (\bibinfo{year}{2003}).

\bibitem[{\citenamefont{Schunck and Egido}(2008)}]{Schunck2008}
\bibinfo{author}{\bibfnamefont{N.}~\bibnamefont{Schunck}} \bibnamefont{and}
  \bibinfo{author}{\bibfnamefont{J.~L.} \bibnamefont{Egido}},
  \bibinfo{journal}{Phys. Rev. C} \textbf{\bibinfo{volume}{77}},
  \bibinfo{pages}{011301} (\bibinfo{year}{2008}).

\bibitem[{\citenamefont{Nik{\v s}i{\'c} et~al.}(2014)\citenamefont{Nik{\v
  s}i{\'c}, Paar, Vretenar, and Ring}}]{Niksic2014}
\bibinfo{author}{\bibfnamefont{T.}~\bibnamefont{Nik{\v s}i{\'c}}},
  \bibinfo{author}{\bibfnamefont{N.}~\bibnamefont{Paar}},
  \bibinfo{author}{\bibfnamefont{D.}~\bibnamefont{Vretenar}}, \bibnamefont{and}
  \bibinfo{author}{\bibfnamefont{P.}~\bibnamefont{Ring}},
  \bibinfo{journal}{Comput. phys. Comm.} \textbf{\bibinfo{volume}{185}},
  \bibinfo{pages}{1808} (\bibinfo{year}{2014}).

\bibitem[{\citenamefont{Bouyssy et~al.}(1987)\citenamefont{Bouyssy, Mathiot,
  Van~Giai, and Marcos}}]{Bouyssy1987}
\bibinfo{author}{\bibfnamefont{A.}~\bibnamefont{Bouyssy}},
  \bibinfo{author}{\bibfnamefont{J.-F.} \bibnamefont{Mathiot}},
  \bibinfo{author}{\bibfnamefont{N.}~\bibnamefont{Van~Giai}}, \bibnamefont{and}
  \bibinfo{author}{\bibfnamefont{S.}~\bibnamefont{Marcos}},
  \bibinfo{journal}{Phys. Rev. C} \textbf{\bibinfo{volume}{36}},
  \bibinfo{pages}{380} (\bibinfo{year}{1987}).

\bibitem[{\citenamefont{Ring}(1996)}]{Ring1996}
\bibinfo{author}{\bibfnamefont{P.}~\bibnamefont{Ring}}, \bibinfo{journal}{Prog.
  Part. Nucl. Phys} \textbf{\bibinfo{volume}{37}}, \bibinfo{pages}{193}
  (\bibinfo{year}{1996}).

\bibitem[{\citenamefont{Bender et~al.}(2003)\citenamefont{Bender, Heenen, and
  Reinhard}}]{Bender2003}
\bibinfo{author}{\bibfnamefont{M.}~\bibnamefont{Bender}},
  \bibinfo{author}{\bibfnamefont{P.-H.} \bibnamefont{Heenen}},
  \bibnamefont{and} \bibinfo{author}{\bibfnamefont{P.-G.}
  \bibnamefont{Reinhard}}, \bibinfo{journal}{Rev. Mod. Phys.}
  \textbf{\bibinfo{volume}{75}}, \bibinfo{pages}{121} (\bibinfo{year}{2003}).

\bibitem[{\citenamefont{Vretenar et~al.}(2005)\citenamefont{Vretenar,
  Afanasjev, Lalazissis, and Ring}}]{Vretenara2005}
\bibinfo{author}{\bibfnamefont{D.}~\bibnamefont{Vretenar}},
  \bibinfo{author}{\bibfnamefont{A.~V.} \bibnamefont{Afanasjev}},
  \bibinfo{author}{\bibfnamefont{G.~A.} \bibnamefont{Lalazissis}},
  \bibnamefont{and} \bibinfo{author}{\bibfnamefont{P.}~\bibnamefont{Ring}},
  \bibinfo{journal}{Phys. Rep.} \textbf{\bibinfo{volume}{409}},
  \bibinfo{pages}{101} (\bibinfo{year}{2005}).

\bibitem[{\citenamefont{Meng et~al.}(2006)\citenamefont{Meng, Toki, Zhou,
  Zhang, Long, and Geng}}]{Meng2006}
\bibinfo{author}{\bibfnamefont{J.}~\bibnamefont{Meng}},
  \bibinfo{author}{\bibfnamefont{H.}~\bibnamefont{Toki}},
  \bibinfo{author}{\bibfnamefont{S.~G.} \bibnamefont{Zhou}},
  \bibinfo{author}{\bibfnamefont{S.~Q.} \bibnamefont{Zhang}},
  \bibinfo{author}{\bibfnamefont{W.~H.} \bibnamefont{Long}}, \bibnamefont{and}
  \bibinfo{author}{\bibfnamefont{L.~S.} \bibnamefont{Geng}},
  \bibinfo{journal}{Prog. Part. Nucl. Phys} \textbf{\bibinfo{volume}{57}},
  \bibinfo{pages}{470} (\bibinfo{year}{2006}).

\bibitem[{\citenamefont{Long et~al.}(2006)\citenamefont{Long, Van~Giai, and
  Meng}}]{Long2006}
\bibinfo{author}{\bibfnamefont{W.~H.} \bibnamefont{Long}},
  \bibinfo{author}{\bibfnamefont{N.}~\bibnamefont{Van~Giai}}, \bibnamefont{and}
  \bibinfo{author}{\bibfnamefont{J.}~\bibnamefont{Meng}},
  \bibinfo{journal}{Phys. Lett. B} \textbf{\bibinfo{volume}{640}},
  \bibinfo{pages}{150} (\bibinfo{year}{2006}).

\bibitem[{\citenamefont{Berger et~al.}(1984)\citenamefont{Berger, Girod, and
  Gogny}}]{Berger1984}
\bibinfo{author}{\bibfnamefont{J.~F.} \bibnamefont{Berger}},
  \bibinfo{author}{\bibfnamefont{M.}~\bibnamefont{Girod}}, \bibnamefont{and}
  \bibinfo{author}{\bibfnamefont{D.}~\bibnamefont{Gogny}},
  \bibinfo{journal}{Nucl. Phys. A} \textbf{\bibinfo{volume}{428}},
  \bibinfo{pages}{23} (\bibinfo{year}{1984}).

\bibitem[{\citenamefont{Tajima et~al.}(1993)\citenamefont{Tajima, Bonche,
  Flocard, Heenen, and Weiss}}]{Tajima1993}
\bibinfo{author}{\bibfnamefont{N.}~\bibnamefont{Tajima}},
  \bibinfo{author}{\bibfnamefont{P.}~\bibnamefont{Bonche}},
  \bibinfo{author}{\bibfnamefont{H.}~\bibnamefont{Flocard}},
  \bibinfo{author}{\bibfnamefont{P.-H.} \bibnamefont{Heenen}},
  \bibnamefont{and} \bibinfo{author}{\bibfnamefont{M.~S.} \bibnamefont{Weiss}},
  \bibinfo{journal}{Nucl. Phys. A} \textbf{\bibinfo{volume}{551}},
  \bibinfo{pages}{434} (\bibinfo{year}{1993}).

\bibitem[{\citenamefont{Long et~al.}(2007)\citenamefont{Long, Sagawa, Van~Giai,
  and Meng}}]{Long2007}
\bibinfo{author}{\bibfnamefont{W.}~\bibnamefont{Long}},
  \bibinfo{author}{\bibfnamefont{H.}~\bibnamefont{Sagawa}},
  \bibinfo{author}{\bibfnamefont{N.}~\bibnamefont{Van~Giai}}, \bibnamefont{and}
  \bibinfo{author}{\bibfnamefont{J.}~\bibnamefont{Meng}},
  \bibinfo{journal}{Phys. Rev. C} \textbf{\bibinfo{volume}{76}},
  \bibinfo{pages}{034314} (\bibinfo{year}{2007}).

\bibitem[{\citenamefont{Kucharek and Ring}(1991)}]{Kucharek1991}
\bibinfo{author}{\bibfnamefont{H.}~\bibnamefont{Kucharek}} \bibnamefont{and}
  \bibinfo{author}{\bibfnamefont{P.}~\bibnamefont{Ring}}, \bibinfo{journal}{Z.
  Phys. A} \textbf{\bibinfo{volume}{339}}, \bibinfo{pages}{23}
  (\bibinfo{year}{1991}).

\bibitem[{\citenamefont{Long et~al.}(2010)\citenamefont{Long, Ring, Van~Giai,
  and Meng}}]{Long2010}
\bibinfo{author}{\bibfnamefont{W.~H.} \bibnamefont{Long}},
  \bibinfo{author}{\bibfnamefont{P.}~\bibnamefont{Ring}},
  \bibinfo{author}{\bibfnamefont{N.}~\bibnamefont{Van~Giai}}, \bibnamefont{and}
  \bibinfo{author}{\bibfnamefont{J.}~\bibnamefont{Meng}},
  \bibinfo{journal}{Phys. Rev. C} \textbf{\bibinfo{volume}{81}},
  \bibinfo{pages}{024308} (\bibinfo{year}{2010}).

\bibitem[{\citenamefont{Ebran et~al.}(2011)\citenamefont{Ebran, Khan, Pe\~na
  Arteaga, and Vretenar}}]{Ebran2011}
\bibinfo{author}{\bibfnamefont{J.-P.} \bibnamefont{Ebran}},
  \bibinfo{author}{\bibfnamefont{E.}~\bibnamefont{Khan}},
  \bibinfo{author}{\bibfnamefont{D.}~\bibnamefont{Pe\~na Arteaga}},
  \bibnamefont{and} \bibinfo{author}{\bibfnamefont{D.}~\bibnamefont{Vretenar}},
  \bibinfo{journal}{Phys. Rev. C} \textbf{\bibinfo{volume}{83}},
  \bibinfo{pages}{064323} (\bibinfo{year}{2011}).

\bibitem[{\citenamefont{Bonche et~al.}(1984)\citenamefont{Bonche, Levit, and
  Vautherin}}]{Bonche1984}
\bibinfo{author}{\bibfnamefont{P.}~\bibnamefont{Bonche}},
  \bibinfo{author}{\bibfnamefont{S.}~\bibnamefont{Levit}}, \bibnamefont{and}
  \bibinfo{author}{\bibfnamefont{D.}~\bibnamefont{Vautherin}},
  \bibinfo{journal}{Nucl. Phys. A} \textbf{\bibinfo{volume}{427}},
  \bibinfo{pages}{278} (\bibinfo{year}{1984}).

\bibitem[{\citenamefont{Bender et~al.}(2000)\citenamefont{Bender, Rutz,
  Reinhard, and Maruhn}}]{Bender2000}
\bibinfo{author}{\bibfnamefont{M.}~\bibnamefont{Bender}},
  \bibinfo{author}{\bibfnamefont{K.}~\bibnamefont{Rutz}},
  \bibinfo{author}{\bibfnamefont{P.-G.} \bibnamefont{Reinhard}},
  \bibnamefont{and} \bibinfo{author}{\bibfnamefont{J.~A.}
  \bibnamefont{Maruhn}}, \bibinfo{journal}{Eur. Phys. J. A}
  \textbf{\bibinfo{volume}{7}}, \bibinfo{pages}{467} (\bibinfo{year}{2000}).

\bibitem[{\citenamefont{Long et~al.}(2008)\citenamefont{Long, Sagawa, Meng, and
  Van~Giai}}]{Long2008}
\bibinfo{author}{\bibfnamefont{W.}~\bibnamefont{Long}},
  \bibinfo{author}{\bibfnamefont{H.}~\bibnamefont{Sagawa}},
  \bibinfo{author}{\bibfnamefont{J.}~\bibnamefont{Meng}}, \bibnamefont{and}
  \bibinfo{author}{\bibfnamefont{N.}~\bibnamefont{Van~Giai}},
  \bibinfo{journal}{Europhys. Lett} \textbf{\bibinfo{volume}{82}},
  \bibinfo{pages}{12001} (\bibinfo{year}{2008}).

\bibitem[{\citenamefont{Lalazissis et~al.}(2005)\citenamefont{Lalazissis,
  Nik{\v s}i{\'c}, Vretenar, and Ring}}]{Lalazissis2005}
\bibinfo{author}{\bibfnamefont{G.~A.} \bibnamefont{Lalazissis}},
  \bibinfo{author}{\bibfnamefont{T.}~\bibnamefont{Nik{\v s}i{\'c}}},
  \bibinfo{author}{\bibfnamefont{D.}~\bibnamefont{Vretenar}}, \bibnamefont{and}
  \bibinfo{author}{\bibfnamefont{P.}~\bibnamefont{Ring}},
  \bibinfo{journal}{Phys. Rev. C} \textbf{\bibinfo{volume}{71}},
  \bibinfo{pages}{024312} (\bibinfo{year}{2005}).

\bibitem[{\citenamefont{Long et~al.}(2004)\citenamefont{Long, Meng, Van~Giai,
  and Zhou}}]{Long2004}
\bibinfo{author}{\bibfnamefont{W.}~\bibnamefont{Long}},
  \bibinfo{author}{\bibfnamefont{J.}~\bibnamefont{Meng}},
  \bibinfo{author}{\bibfnamefont{N.}~\bibnamefont{Van~Giai}}, \bibnamefont{and}
  \bibinfo{author}{\bibfnamefont{S.-G.} \bibnamefont{Zhou}},
  \bibinfo{journal}{Phys. Rev. C} \textbf{\bibinfo{volume}{69}},
  \bibinfo{pages}{034319} (\bibinfo{year}{2004}).

\bibitem[{\citenamefont{Lalazissis et~al.}(2009)\citenamefont{Lalazissis,
  Karatzikos, Fossion, Pena~Arteaga, Afanasjev, and Ring}}]{Lalazissis2009}
\bibinfo{author}{\bibfnamefont{G.}~\bibnamefont{Lalazissis}},
  \bibinfo{author}{\bibfnamefont{S.}~\bibnamefont{Karatzikos}},
  \bibinfo{author}{\bibfnamefont{R.}~\bibnamefont{Fossion}},
  \bibinfo{author}{\bibfnamefont{D.}~\bibnamefont{Pena~Arteaga}},
  \bibinfo{author}{\bibfnamefont{A.}~\bibnamefont{Afanasjev}},
  \bibnamefont{and} \bibinfo{author}{\bibfnamefont{P.}~\bibnamefont{Ring}},
  \bibinfo{journal}{Phys. lett. B} \textbf{\bibinfo{volume}{671}},
  \bibinfo{pages}{36} (\bibinfo{year}{2009}).

\bibitem[{\citenamefont{Wang et~al.}(2013)\citenamefont{Wang, Sun, Dong, and
  Long}}]{Wang2013}
\bibinfo{author}{\bibfnamefont{L.~J.} \bibnamefont{Wang}},
  \bibinfo{author}{\bibfnamefont{B.~Y.} \bibnamefont{Sun}},
  \bibinfo{author}{\bibfnamefont{J.~M.} \bibnamefont{Dong}}, \bibnamefont{and}
  \bibinfo{author}{\bibfnamefont{W.~H.} \bibnamefont{Long}},
  \bibinfo{journal}{Phys. Rev. C} \textbf{\bibinfo{volume}{87}},
  \bibinfo{pages}{054331} (\bibinfo{year}{2013}).

\bibitem[{\citenamefont{Agbemava et~al.}(2014)\citenamefont{Agbemava,
  Afanasjev, Ray, and Ring}}]{Agbemava2014}
\bibinfo{author}{\bibfnamefont{S.~E.} \bibnamefont{Agbemava}},
  \bibinfo{author}{\bibfnamefont{A.~V.} \bibnamefont{Afanasjev}},
  \bibinfo{author}{\bibfnamefont{D.}~\bibnamefont{Ray}}, \bibnamefont{and}
  \bibinfo{author}{\bibfnamefont{P.}~\bibnamefont{Ring}},
  \bibinfo{journal}{Phys. Rev. C} \textbf{\bibinfo{volume}{89}},
  \bibinfo{pages}{054320} (\bibinfo{year}{2014}).

\bibitem[{\citenamefont{Jaminon and Mahaux}(1989)}]{Jaminon1989}
\bibinfo{author}{\bibfnamefont{M.}~\bibnamefont{Jaminon}} \bibnamefont{and}
  \bibinfo{author}{\bibfnamefont{C.}~\bibnamefont{Mahaux}},
  \bibinfo{journal}{Phys. Rev. C} \textbf{\bibinfo{volume}{40}},
  \bibinfo{pages}{354} (\bibinfo{year}{1989}).

\bibitem[{\citenamefont{Lombardo}(1999)}]{Baldo1999}
\bibinfo{author}{\bibfnamefont{U.}~\bibnamefont{Lombardo}},
  \emph{\bibinfo{title}{Nuclear Methods and the Nuclear Equation of State}}
  (\bibinfo{publisher}{World Scientific}, \bibinfo{address}{Singapore},
  \bibinfo{year}{1999}), chap. \bibinfo{chapter}{Superfluidity in Nuclear
  Matter}.

\bibitem[{\citenamefont{Dobaczewski et~al.}(1996)\citenamefont{Dobaczewski,
  Nazarewicz, Werner, Berger, Chinn, and Decharg\'e}}]{Dobaczewski1996}
\bibinfo{author}{\bibfnamefont{J.}~\bibnamefont{Dobaczewski}},
  \bibinfo{author}{\bibfnamefont{W.}~\bibnamefont{Nazarewicz}},
  \bibinfo{author}{\bibfnamefont{T.~R.} \bibnamefont{Werner}},
  \bibinfo{author}{\bibfnamefont{J.~F.} \bibnamefont{Berger}},
  \bibinfo{author}{\bibfnamefont{C.~R.} \bibnamefont{Chinn}}, \bibnamefont{and}
  \bibinfo{author}{\bibfnamefont{J.}~\bibnamefont{Decharg\'e}},
  \bibinfo{journal}{Phys. Rev. C} \textbf{\bibinfo{volume}{53}},
  \bibinfo{pages}{2809} (\bibinfo{year}{1996}).

\bibitem[{\citenamefont{Goriely}(1996)}]{Goriely1996}
\bibinfo{author}{\bibfnamefont{S.}~\bibnamefont{Goriely}},
  \bibinfo{journal}{Nucl. Phys. A} \textbf{\bibinfo{volume}{605}},
  \bibinfo{pages}{28} (\bibinfo{year}{1996}).

\bibitem[{\citenamefont{Bertsch et~al.}(2009)\citenamefont{Bertsch, Bertulani,
  Nazarewicz, Schunck, and Stoitsov}}]{Bertsch2009}
\bibinfo{author}{\bibfnamefont{G.~F.} \bibnamefont{Bertsch}},
  \bibinfo{author}{\bibfnamefont{C.~A.} \bibnamefont{Bertulani}},
  \bibinfo{author}{\bibfnamefont{W.}~\bibnamefont{Nazarewicz}},
  \bibinfo{author}{\bibfnamefont{N.}~\bibnamefont{Schunck}}, \bibnamefont{and}
  \bibinfo{author}{\bibfnamefont{M.~V.} \bibnamefont{Stoitsov}},
  \bibinfo{journal}{Phys. Rev. C} \textbf{\bibinfo{volume}{79}},
  \bibinfo{pages}{034306} (\bibinfo{year}{2009}).

\bibitem[{\citenamefont{Y\"uksel et~al.}(2014)\citenamefont{Y\"uksel, Khan,
  Bozkurt, and Col\'o}}]{Yuksel2014}
\bibinfo{author}{\bibfnamefont{E.}~\bibnamefont{Y\"uksel}},
  \bibinfo{author}{\bibfnamefont{E.}~\bibnamefont{Khan}},
  \bibinfo{author}{\bibfnamefont{K.}~\bibnamefont{Bozkurt}}, \bibnamefont{and}
  \bibinfo{author}{\bibfnamefont{G.}~\bibnamefont{Col\'o}},
  \bibinfo{journal}{Eur. Phys. J. A} \textbf{\bibinfo{volume}{50}},
  \bibinfo{pages}{160} (\bibinfo{year}{2014}).

\bibitem[{\citenamefont{Nik{\v s}i{\'c} et~al.}(2002)\citenamefont{Nik{\v
  s}i{\'c}, Vretenar, Finelli, and Ring}}]{Niksic2002}
\bibinfo{author}{\bibfnamefont{T.}~\bibnamefont{Nik{\v s}i{\'c}}},
  \bibinfo{author}{\bibfnamefont{D.}~\bibnamefont{Vretenar}},
  \bibinfo{author}{\bibfnamefont{P.}~\bibnamefont{Finelli}}, \bibnamefont{and}
  \bibinfo{author}{\bibfnamefont{P.}~\bibnamefont{Ring}},
  \bibinfo{journal}{Phys. Rev. C} \textbf{\bibinfo{volume}{66}},
  \bibinfo{pages}{024306} (\bibinfo{year}{2002}).

\bibitem[{\citenamefont{Roca-Maza et~al.}(2011)\citenamefont{Roca-Maza,
  Vi\~nas, Centelles, Ring, and Schuck}}]{Roca2011}
\bibinfo{author}{\bibfnamefont{X.}~\bibnamefont{Roca-Maza}},
  \bibinfo{author}{\bibfnamefont{X.}~\bibnamefont{Vi\~nas}},
  \bibinfo{author}{\bibfnamefont{M.}~\bibnamefont{Centelles}},
  \bibinfo{author}{\bibfnamefont{P.}~\bibnamefont{Ring}}, \bibnamefont{and}
  \bibinfo{author}{\bibfnamefont{P.}~\bibnamefont{Schuck}},
  \bibinfo{journal}{Phys. Rev. C} \textbf{\bibinfo{volume}{84}},
  \bibinfo{pages}{054309} (\bibinfo{year}{2011}).

\bibitem[{\citenamefont{Typel and Wolter}(1999)}]{Typel1999}
\bibinfo{author}{\bibfnamefont{S.}~\bibnamefont{Typel}} \bibnamefont{and}
  \bibinfo{author}{\bibfnamefont{H.~H.} \bibnamefont{Wolter}},
  \bibinfo{journal}{Nucl. Phys. A} \textbf{\bibinfo{volume}{656}},
  \bibinfo{pages}{331} (\bibinfo{year}{1999}).

\bibitem[{\citenamefont{Lalazissis et~al.}(1997)\citenamefont{Lalazissis,
  K\"onig, and Ring}}]{Lalazissis1997}
\bibinfo{author}{\bibfnamefont{G.~A.} \bibnamefont{Lalazissis}},
  \bibinfo{author}{\bibfnamefont{J.}~\bibnamefont{K\"onig}}, \bibnamefont{and}
  \bibinfo{author}{\bibfnamefont{P.}~\bibnamefont{Ring}},
  \bibinfo{journal}{Phys. Rev. C} \textbf{\bibinfo{volume}{55}},
  \bibinfo{pages}{540} (\bibinfo{year}{1997}).

\bibitem[{\citenamefont{Sugahara and Toki}(1994)}]{Sugahara1994}
\bibinfo{author}{\bibfnamefont{Y.}~\bibnamefont{Sugahara}} \bibnamefont{and}
  \bibinfo{author}{\bibfnamefont{H.}~\bibnamefont{Toki}},
  \bibinfo{journal}{Nucl. Phys. A} \textbf{\bibinfo{volume}{579}},
  \bibinfo{pages}{557} (\bibinfo{year}{1994}).

\bibitem[{\citenamefont{Tian et~al.}(2009)\citenamefont{Tian, Ma, and
  Ring}}]{Tian2009}
\bibinfo{author}{\bibfnamefont{Y.}~\bibnamefont{Tian}},
  \bibinfo{author}{\bibfnamefont{Z.~Y.} \bibnamefont{Ma}}, \bibnamefont{and}
  \bibinfo{author}{\bibfnamefont{P.}~\bibnamefont{Ring}},
  \bibinfo{journal}{Phys. Lett. B} \textbf{\bibinfo{volume}{676}},
  \bibinfo{pages}{44} (\bibinfo{year}{2009}).

\bibitem[{\citenamefont{Zhao et~al.}(2010)\citenamefont{Zhao, Li, Yao, and
  Meng}}]{Zhao2010}
\bibinfo{author}{\bibfnamefont{P.~W.} \bibnamefont{Zhao}},
  \bibinfo{author}{\bibfnamefont{Z.~P.} \bibnamefont{Li}},
  \bibinfo{author}{\bibfnamefont{J.~M.} \bibnamefont{Yao}}, \bibnamefont{and}
  \bibinfo{author}{\bibfnamefont{J.}~\bibnamefont{Meng}},
  \bibinfo{journal}{Phys. Rev. C} \textbf{\bibinfo{volume}{82}},
  \bibinfo{pages}{054319} (\bibinfo{year}{2010}).

\bibitem[{\citenamefont{Mahan}(2000)}]{Mahan2000}
\bibinfo{author}{\bibfnamefont{G.~D.} \bibnamefont{Mahan}},
  \emph{\bibinfo{title}{Many-Particle Physics}} (\bibinfo{publisher}{Plenum
  Press}, \bibinfo{address}{New York}, \bibinfo{year}{2000}).

\bibitem[{\citenamefont{Erler et~al.}(2012)\citenamefont{Erler, Birge,
  Kortelainen, Nazarewicz, Olsen, Perhac, and Stoitsov}}]{Erler2012}
\bibinfo{author}{\bibfnamefont{J.}~\bibnamefont{Erler}},
  \bibinfo{author}{\bibfnamefont{N.}~\bibnamefont{Birge}},
  \bibinfo{author}{\bibfnamefont{M.}~\bibnamefont{Kortelainen}},
  \bibinfo{author}{\bibfnamefont{W.}~\bibnamefont{Nazarewicz}},
  \bibinfo{author}{\bibfnamefont{E.}~\bibnamefont{Olsen}},
  \bibinfo{author}{\bibfnamefont{A.~M.} \bibnamefont{Perhac}},
  \bibnamefont{and} \bibinfo{author}{\bibfnamefont{M.}~\bibnamefont{Stoitsov}},
  \bibinfo{journal}{Nature} \textbf{\bibinfo{volume}{486}},
  \bibinfo{pages}{509} (\bibinfo{year}{2012}).

\bibitem[{\citenamefont{Erler et~al.}(2013)\citenamefont{Erler, Horowitz,
  Nazarewicz, Rafalski, and Reinhard}}]{Erler2013}
\bibinfo{author}{\bibfnamefont{J.}~\bibnamefont{Erler}},
  \bibinfo{author}{\bibfnamefont{C.~J.} \bibnamefont{Horowitz}},
  \bibinfo{author}{\bibfnamefont{W.}~\bibnamefont{Nazarewicz}},
  \bibinfo{author}{\bibfnamefont{M.}~\bibnamefont{Rafalski}}, \bibnamefont{and}
  \bibinfo{author}{\bibfnamefont{P.-G.} \bibnamefont{Reinhard}},
  \bibinfo{journal}{Phys. Rev. C} \textbf{\bibinfo{volume}{87}},
  \bibinfo{pages}{044320} (\bibinfo{year}{2013}).

\bibitem[{\citenamefont{Hilaire and Girod}(2007)}]{Hilaire2007}
\bibinfo{author}{\bibfnamefont{S.}~\bibnamefont{Hilaire}} \bibnamefont{and}
  \bibinfo{author}{\bibfnamefont{M.}~\bibnamefont{Girod}},
  \bibinfo{journal}{Eur. Phys. J. A} \textbf{\bibinfo{volume}{33}},
  \bibinfo{pages}{237} (\bibinfo{year}{2007}).

\bibitem[{\citenamefont{Delaroche et~al.}(2010)\citenamefont{Delaroche, Girod,
  Libert, Goutte, Hilaire, P\'eru, Pillet, and Bertsch}}]{Delaroche2010}
\bibinfo{author}{\bibfnamefont{J.~P.} \bibnamefont{Delaroche}},
  \bibinfo{author}{\bibfnamefont{M.}~\bibnamefont{Girod}},
  \bibinfo{author}{\bibfnamefont{J.}~\bibnamefont{Libert}},
  \bibinfo{author}{\bibfnamefont{H.}~\bibnamefont{Goutte}},
  \bibinfo{author}{\bibfnamefont{S.}~\bibnamefont{Hilaire}},
  \bibinfo{author}{\bibfnamefont{S.}~\bibnamefont{P\'eru}},
  \bibinfo{author}{\bibfnamefont{N.}~\bibnamefont{Pillet}}, \bibnamefont{and}
  \bibinfo{author}{\bibfnamefont{G.~F.} \bibnamefont{Bertsch}},
  \bibinfo{journal}{Phys. Rev. C} \textbf{\bibinfo{volume}{81}},
  \bibinfo{pages}{014303} (\bibinfo{year}{2010}).

\bibitem[{\citenamefont{Afanasjev et~al.}(2013)\citenamefont{Afanasjev,
  Agbemava, Ray, and Ring}}]{Afanasjev2013}
\bibinfo{author}{\bibfnamefont{A.~V.} \bibnamefont{Afanasjev}},
  \bibinfo{author}{\bibfnamefont{S.~E.} \bibnamefont{Agbemava}},
  \bibinfo{author}{\bibfnamefont{D.}~\bibnamefont{Ray}}, \bibnamefont{and}
  \bibinfo{author}{\bibfnamefont{P.}~\bibnamefont{Ring}},
  \bibinfo{journal}{Phys. Lett. B} \textbf{\bibinfo{volume}{726}},
  \bibinfo{pages}{680} (\bibinfo{year}{2013}).

\bibitem[{\citenamefont{Long et~al.}(2009)\citenamefont{Long, Nakatsukasa,
  Sagawa, Meng, Nakada, and Zhang}}]{Long2009}
\bibinfo{author}{\bibfnamefont{W.}~\bibnamefont{Long}},
  \bibinfo{author}{\bibfnamefont{T.}~\bibnamefont{Nakatsukasa}},
  \bibinfo{author}{\bibfnamefont{H.}~\bibnamefont{Sagawa}},
  \bibinfo{author}{\bibfnamefont{J.}~\bibnamefont{Meng}},
  \bibinfo{author}{\bibfnamefont{H.}~\bibnamefont{Nakada}}, \bibnamefont{and}
  \bibinfo{author}{\bibfnamefont{Y.}~\bibnamefont{Zhang}},
  \bibinfo{journal}{Phys. Lett. B} \textbf{\bibinfo{volume}{680}},
  \bibinfo{pages}{428} (\bibinfo{year}{2009}).

\bibitem[{\citenamefont{Li et~al.}(2014)\citenamefont{Li, Long, Margueron, and
  Van~Giai}}]{Li2014}
\bibinfo{author}{\bibfnamefont{J.~J.} \bibnamefont{Li}},
  \bibinfo{author}{\bibfnamefont{W.~H.} \bibnamefont{Long}},
  \bibinfo{author}{\bibfnamefont{J.}~\bibnamefont{Margueron}},
  \bibnamefont{and} \bibinfo{author}{\bibfnamefont{N.}~\bibnamefont{Van~Giai}},
  \bibinfo{journal}{Phys. Lett. B} \textbf{\bibinfo{volume}{732}},
  \bibinfo{pages}{169} (\bibinfo{year}{2014}).

\bibitem[{\citenamefont{Sorlin et~al.}(2002)\citenamefont{Sorlin, Leenhardt,
  Donzaud, Duprat, Azaiez, Nowacki, Grawe, Dombr\'adi, Amorini, Astier
  et~al.}}]{Sorlin2002}
\bibinfo{author}{\bibfnamefont{O.}~\bibnamefont{Sorlin}},
  \bibinfo{author}{\bibfnamefont{S.}~\bibnamefont{Leenhardt}},
  \bibinfo{author}{\bibfnamefont{C.}~\bibnamefont{Donzaud}},
  \bibinfo{author}{\bibfnamefont{J.}~\bibnamefont{Duprat}},
  \bibinfo{author}{\bibfnamefont{F.}~\bibnamefont{Azaiez}},
  \bibinfo{author}{\bibfnamefont{F.}~\bibnamefont{Nowacki}},
  \bibinfo{author}{\bibfnamefont{H.}~\bibnamefont{Grawe}},
  \bibinfo{author}{\bibfnamefont{Z.}~\bibnamefont{Dombr\'adi}},
  \bibinfo{author}{\bibfnamefont{F.}~\bibnamefont{Amorini}},
  \bibinfo{author}{\bibfnamefont{A.}~\bibnamefont{Astier}},
  \bibnamefont{et~al.}, \bibinfo{journal}{Phys. Rev. Lett.}
  \textbf{\bibinfo{volume}{88}}, \bibinfo{pages}{092501}
  (\bibinfo{year}{2002}).

\bibitem[{\citenamefont{Geng et~al.}(2006)\citenamefont{Geng, Meng, Toki, Long,
  and Shen}}]{Geng2006}
\bibinfo{author}{\bibfnamefont{L.-S.} \bibnamefont{Geng}},
  \bibinfo{author}{\bibfnamefont{J.}~\bibnamefont{Meng}},
  \bibinfo{author}{\bibfnamefont{H.}~\bibnamefont{Toki}},
  \bibinfo{author}{\bibfnamefont{W.-H.} \bibnamefont{Long}}, \bibnamefont{and}
  \bibinfo{author}{\bibfnamefont{G.}~\bibnamefont{Shen}},
  \bibinfo{journal}{Chin. Phys. Lett} \textbf{\bibinfo{volume}{23}},
  \bibinfo{pages}{1139} (\bibinfo{year}{2006}).

\bibitem[{\citenamefont{Pastore}(2012)}]{Pastore2012}
\bibinfo{author}{\bibfnamefont{A.}~\bibnamefont{Pastore}},
  \bibinfo{journal}{Phys. Rev. C} \textbf{\bibinfo{volume}{86}},
  \bibinfo{pages}{065802} (\bibinfo{year}{2012}).

\bibitem[{\citenamefont{Dobaczewski et~al.}(1994)\citenamefont{Dobaczewski,
  Hamamoto, Nazarewicz, and Sheikh}}]{Dobaczewski1994}
\bibinfo{author}{\bibfnamefont{J.}~\bibnamefont{Dobaczewski}},
  \bibinfo{author}{\bibfnamefont{I.}~\bibnamefont{Hamamoto}},
  \bibinfo{author}{\bibfnamefont{W.}~\bibnamefont{Nazarewicz}},
  \bibnamefont{and} \bibinfo{author}{\bibfnamefont{J.~A.}
  \bibnamefont{Sheikh}}, \bibinfo{journal}{Phys. Rev. Lett.}
  \textbf{\bibinfo{volume}{72}}, \bibinfo{pages}{981} (\bibinfo{year}{1994}).

\bibitem[{\citenamefont{Kleban et~al.}(2002)\citenamefont{Kleban,
  Nerlo-Pomorska, Berger, Decharg\'e, Girod, and Hilaire}}]{Kleban2002}
\bibinfo{author}{\bibfnamefont{M.}~\bibnamefont{Kleban}},
  \bibinfo{author}{\bibfnamefont{B.}~\bibnamefont{Nerlo-Pomorska}},
  \bibinfo{author}{\bibfnamefont{J.~F.} \bibnamefont{Berger}},
  \bibinfo{author}{\bibfnamefont{J.}~\bibnamefont{Decharg\'e}},
  \bibinfo{author}{\bibfnamefont{M.}~\bibnamefont{Girod}}, \bibnamefont{and}
  \bibinfo{author}{\bibfnamefont{S.}~\bibnamefont{Hilaire}},
  \bibinfo{journal}{Phys. Rev. C} \textbf{\bibinfo{volume}{65}},
  \bibinfo{pages}{024309} (\bibinfo{year}{2002}).

\bibitem[{\citenamefont{Yamagami et~al.}(2012)\citenamefont{Yamagami,
  Margueron, Sagawa, and Hagino}}]{Yamagami2012}
\bibinfo{author}{\bibfnamefont{M.}~\bibnamefont{Yamagami}},
  \bibinfo{author}{\bibfnamefont{J.}~\bibnamefont{Margueron}},
  \bibinfo{author}{\bibfnamefont{H.}~\bibnamefont{Sagawa}}, \bibnamefont{and}
  \bibinfo{author}{\bibfnamefont{K.}~\bibnamefont{Hagino}},
  \bibinfo{journal}{Phys. Rev. C} \textbf{\bibinfo{volume}{86}},
  \bibinfo{pages}{034333} (\bibinfo{year}{2012}).

\bibitem[{\citenamefont{Meng et~al.}(2002)\citenamefont{Meng, Toki, Zeng,
  Zhang, and Zhou}}]{Meng2002}
\bibinfo{author}{\bibfnamefont{J.}~\bibnamefont{Meng}},
  \bibinfo{author}{\bibfnamefont{H.}~\bibnamefont{Toki}},
  \bibinfo{author}{\bibfnamefont{J.~Y.} \bibnamefont{Zeng}},
  \bibinfo{author}{\bibfnamefont{S.~Q.} \bibnamefont{Zhang}}, \bibnamefont{and}
  \bibinfo{author}{\bibfnamefont{S.-G.} \bibnamefont{Zhou}},
  \bibinfo{journal}{Phys. Rev. C} \textbf{\bibinfo{volume}{65}},
  \bibinfo{pages}{041302} (\bibinfo{year}{2002}).

\bibitem[{\citenamefont{Bennaceur et~al.}(1999)\citenamefont{Bennaceur,
  Dobaczewski, and P{\l}oszajczak}}]{Bennaceur1999}
\bibinfo{author}{\bibfnamefont{K.}~\bibnamefont{Bennaceur}},
  \bibinfo{author}{\bibfnamefont{J.}~\bibnamefont{Dobaczewski}},
  \bibnamefont{and}
  \bibinfo{author}{\bibfnamefont{M.}~\bibnamefont{P{\l}oszajczak}},
  \bibinfo{journal}{Phys. Rev. C} \textbf{\bibinfo{volume}{60}},
  \bibinfo{pages}{034308} (\bibinfo{year}{1999}).

\bibitem[{\citenamefont{Pastore et~al.}(2015)\citenamefont{Pastore, Chamel, and
  Margueron}}]{Pastore2015}
\bibinfo{author}{\bibfnamefont{A.}~\bibnamefont{Pastore}},
  \bibinfo{author}{\bibfnamefont{N.}~\bibnamefont{Chamel}}, \bibnamefont{and}
  \bibinfo{author}{\bibfnamefont{J.}~\bibnamefont{Margueron}},
  \bibinfo{journal}{Mon. Not. Roy. Astron. Soc.}
  \textbf{\bibinfo{volume}{448}}, \bibinfo{pages}{1887} (\bibinfo{year}{2015}).

\bibitem[{\citenamefont{Gilbert and Cameron}(1965)}]{Gilbert1965}
\bibinfo{author}{\bibfnamefont{A.}~\bibnamefont{Gilbert}} \bibnamefont{and}
  \bibinfo{author}{\bibfnamefont{A.~G.~W.} \bibnamefont{Cameron}},
  \bibinfo{journal}{Can. J. Phys.} \textbf{\bibinfo{volume}{43}},
  \bibinfo{pages}{1446} (\bibinfo{year}{1965}).

\bibitem[{\citenamefont{Schiller et~al.}(2001)\citenamefont{Schiller, Bjerve,
  Guttormsen, Hjorth-Jensen, Ingebretsen, Melby, Messelt, Rekstad, Siem, and
  {\O}deg{\aa}rd}}]{Schiller2001}
\bibinfo{author}{\bibfnamefont{A.}~\bibnamefont{Schiller}},
  \bibinfo{author}{\bibfnamefont{A.}~\bibnamefont{Bjerve}},
  \bibinfo{author}{\bibfnamefont{M.}~\bibnamefont{Guttormsen}},
  \bibinfo{author}{\bibfnamefont{M.}~\bibnamefont{Hjorth-Jensen}},
  \bibinfo{author}{\bibfnamefont{F.}~\bibnamefont{Ingebretsen}},
  \bibinfo{author}{\bibfnamefont{E.}~\bibnamefont{Melby}},
  \bibinfo{author}{\bibfnamefont{S.}~\bibnamefont{Messelt}},
  \bibinfo{author}{\bibfnamefont{J.}~\bibnamefont{Rekstad}},
  \bibinfo{author}{\bibfnamefont{S.}~\bibnamefont{Siem}}, \bibnamefont{and}
  \bibinfo{author}{\bibfnamefont{S.~W.} \bibnamefont{{\O}deg{\aa}rd}},
  \bibinfo{journal}{Phys. Rev. C} \textbf{\bibinfo{volume}{63}},
  \bibinfo{pages}{021306} (\bibinfo{year}{2001}).

\bibitem[{\citenamefont{Tanabe et~al.}(1981)\citenamefont{Tanabe,
  Sugawara-Tanabe, and Mang}}]{Tanabe1981}
\bibinfo{author}{\bibfnamefont{K.}~\bibnamefont{Tanabe}},
  \bibinfo{author}{\bibfnamefont{K.}~\bibnamefont{Sugawara-Tanabe}},
  \bibnamefont{and} \bibinfo{author}{\bibfnamefont{H.~J.} \bibnamefont{Mang}},
  \bibinfo{journal}{Nucl. Phys. A} \textbf{\bibinfo{volume}{357}},
  \bibinfo{pages}{20} (\bibinfo{year}{1981}).

\bibitem[{\citenamefont{Esebbag and Egido}(1993)}]{Esebbag1993}
\bibinfo{author}{\bibfnamefont{C.}~\bibnamefont{Esebbag}} \bibnamefont{and}
  \bibinfo{author}{\bibfnamefont{J.~L.} \bibnamefont{Egido}},
  \bibinfo{journal}{Nucl. Phys. A} \textbf{\bibinfo{volume}{552}},
  \bibinfo{pages}{205} (\bibinfo{year}{1993}).

\bibitem[{\citenamefont{Dinh~Dang and Arima}(2003)}]{Dinh2003}
\bibinfo{author}{\bibfnamefont{N.}~\bibnamefont{Dinh~Dang}} \bibnamefont{and}
  \bibinfo{author}{\bibfnamefont{A.}~\bibnamefont{Arima}},
  \bibinfo{journal}{Phys. Rev. C} \textbf{\bibinfo{volume}{68}},
  \bibinfo{pages}{014318} (\bibinfo{year}{2003}).

\bibitem[{\citenamefont{Pastore et~al.}(2014)\citenamefont{Pastore, Schuck,
  Urban, Vi\~nas, and Margueron}}]{Pastore2014}
\bibinfo{author}{\bibfnamefont{A.}~\bibnamefont{Pastore}},
  \bibinfo{author}{\bibfnamefont{P.}~\bibnamefont{Schuck}},
  \bibinfo{author}{\bibfnamefont{M.}~\bibnamefont{Urban}},
  \bibinfo{author}{\bibfnamefont{X.}~\bibnamefont{Vi\~nas}}, \bibnamefont{and}
  \bibinfo{author}{\bibfnamefont{J.}~\bibnamefont{Margueron}},
  \bibinfo{journal}{Phys. Rev. A} \textbf{\bibinfo{volume}{90}},
  \bibinfo{pages}{043634} (\bibinfo{year}{2014}).

\bibitem[{\citenamefont{Grill et~al.}(2011)\citenamefont{Grill, Margueron, and
  Sandulescu}}]{Grill2011}
\bibinfo{author}{\bibfnamefont{F.}~\bibnamefont{Grill}},
  \bibinfo{author}{\bibfnamefont{J.}~\bibnamefont{Margueron}},
  \bibnamefont{and}
  \bibinfo{author}{\bibfnamefont{N.}~\bibnamefont{Sandulescu}},
  \bibinfo{journal}{Phys. Rev. C} \textbf{\bibinfo{volume}{84}},
  \bibinfo{pages}{065801} (\bibinfo{year}{2011}).

\bibitem[{\citenamefont{Margueron and Sandulescu}(2012)}]{Margueron2012b}
\bibinfo{author}{\bibfnamefont{J.}~\bibnamefont{Margueron}} \bibnamefont{and}
  \bibinfo{author}{\bibfnamefont{N.}~\bibnamefont{Sandulescu}},
  \emph{\bibinfo{title}{Neutron Star Crust}} (\bibinfo{publisher}{World
  Scientific}, \bibinfo{address}{Singapore}, \bibinfo{year}{2012}), chap.
  \bibinfo{chapter}{Pairing Correlations and Thermodynamic Properties of Inner
  Crust Matter}.

\end{thebibliography}
\end{document}